\documentclass[iop, apj]{emulateapj}
\usepackage{graphics}      % standard graphics specifications
\usepackage{amsmath}
\usepackage{amssymb}
\usepackage{morefloats}
\usepackage{natbib}

\makeatletter

\newcommand{\Rmnum}[1]{\expandafter\@slowromancap\romannumeral #1@}
\makeatother

\def\gtrsim{\mathrel{\hbox{\rlap{\hbox{\lower4pt\hbox{$\sim$}}}\hbox{\raise2pt\hbox{$>$}}}}}

\newcommand{\kms}{km~s\ensuremath{^{-1}}}

\newcommand{\mgii}{Mg~{\tiny II}}

\newcommand{\rblr}{\ensuremath{R_{\mathrm{BLR}}}}

\def\lax{{$\mathrel{\hbox{\rlap{\hbox{\lower4pt\hbox{$\sim$}}}\hbox{$<$}}}$}}
\def\gax{{$\mathrel{\hbox{\rlap{\hbox{\lower4pt\hbox{$\sim$}}}\hbox{$>$}}}$}}

\def\ba{\begin{eqnarray}}
\def\ea{\end{eqnarray}}

\shorttitle{Search for Supermassive Black Hole Binaries}
\shortauthors{Ju et al.}

\begin{document}

%%%%%%%%%%%%%%%%%%%%%%%%%%%%%%%%%%%%%%%%%%%%%%%%%%%%%%%%%%%%%%%%%%%%%%%%%%%%%%%%%%%%%

\title{Search for Supermassive Black Hole Binaries in the Sloan Digital Sky Survey Spectroscopic Sample}

\author{Wenhua Ju\altaffilmark{1}, Jenny E. Greene\altaffilmark{1}, Roman R. Rafikov\altaffilmark{1},
Steven J. Bickerton\altaffilmark{2}, \& Carles Badenes\altaffilmark{3}}

\altaffiltext{1}{Dept. of Astrophysical Sciences, Princeton University, Princeton, NJ 08544, USA}
\altaffiltext{2}{Kavli Institute for the Physics and Mathematics of the Universe, Kashiwa, 277-8583, Japan}
\altaffiltext{3}{Department of Physics and Astronomy and Pittsburgh Particle Physics, Astrophysics, and Cosmology Center (PITT-PACC), University of Pittsburgh, 3941 O'Hara Street, Pittsburgh, PA 15260, USA}

%\maketitle

%%%%%%%%%%%%%%%%%%%%%%%%%%%%%%%% Abstract %%%%%%%%%%%%%%%%%%%%%%%%%%%%%%%%%%
\begin{abstract}
  Supermassive black hole (SMBH) binaries are expected in a $\Lambda$
  CDM cosmology given that most (if not all) massive galaxies contain
  a massive black hole at their center. So far, however, direct
  evidence for such binaries has been elusive. We use
  cross-correlation to search for temporal velocity shifts in the
  {\mgii} broad emission lines of $0.36 < z < 2$ quasars
  with multiple observations in the Sloan Digital Sky Survey. For 
  $\sim 10^9$ $M_\odot$ BHs in SMBH binaries, we are sensitive
  to velocity drifts for binary separations of $\sim 0.1$pc
  with orbital periods of $\sim 100$ years.  We find seven candidate
  sub-pc--scale binaries with velocity shifts $> 3.4\sigma \sim 280$
  km~s$^{-1}$, where $\sigma$ is our systematic error. Comparing the
  detectability of SMBH binaries with the number of candidates (N
  $\leq$ 7), we can rule out that most $10^9$ $M_\odot$ BHs exist in
  $\sim$ 0.03-0.2 pc scale binaries, in a scenario where binaries stall
  at sub-pc scales for a Hubble time. We further constrain
  that $\leq$ one-third of quasars host SMBH binaries after
  considering gas-assisted sub-pc evolution of SMBH binaries, although
  this result is very sensitive to the assumed size of the broad line
  region.  We estimate the detectability of SMBH binaries with ongoing
  or next-generation surveys (e.g., BOSS, Subaru Prime Focus
  Spectrograph), taking into account the evolution of the sub-parsec
  binary in circumbinary gas disks. These future observations will
  provide longer time baselines for searches similar to ours and may
  in turn constrain the evolutionary scenarios of SMBH binaries.
\end{abstract}


%%%%%%%%%%%%%%%%%%%%%%%%%%%%%%%%%%%%%%%%%%%%%%%%%%%%%%%%%%%%%%%%%
%%%%%%%%%%%%%%%%%%%%%% Introduction %%%%%%%%%%%%%%%%%%%%%%%%%%%%%%%%%
%%%%%%%%%%%%%%%%%%%%%%%%%%%%%%%%%%%%%%%%%%%%%%%%%%%%%%%%%%%%%%%%%
\section{Introduction}

Supermassive black holes (SMBHs) were first proposed to account for
the energy source of active galactic nuclei (AGNs, e.g.,
\citealt{1965robinson}). Now it is well known that nearly every galaxy
hosts a SMBH in its center \citep{1995kormendy}. At the same time, in
a $\Lambda$CDM cosmology, galaxies grow by hierarchical merging
\citep{1978white}, and galaxy mergers are directly observed
\citep[e.g.,][]{1992schweizer}.  During the merger, the SMBHs from
each progenitor will sink to the center of the merger remnant and
therefore it seems inevitable that the two SMBHs will form a
gravitationally bound pair, known as a SMBH binary 
\citep{1980begelman}.  Merging SMBHs are expected to be the strongest
source of gravitational wave (GW) radiation for future
space-based laser interferometers \citep{2012amaro}. However, the exact
temporal evolution of these binaries, which directly determines their
detection rates, remains uncertain from both a theoretical and
observational perspective. Our goal in this paper is to place
observational constraints on the highly uncertain sub-parsec phase of
SMBH binary evolution.

SMBH binary systems evolve via the extraction of energy and angular
momentum by the stars or gas in the vicinity, which makes the two BHs
spiral in and coalesce.  A possible scenario for SMBH binary evolution
in the relaxed stellar nucleus of two merging galaxies was first
proposed by \citet{1980begelman}:

\begin{enumerate}

\item The two galaxies spiral into the galaxy center via dynamical friction.
  The black holes in the nuclei become gravitationally
  bound when the semi-major axis of their orbit reaches $r_B \sim 32.2
  \mbox{ pc} (M_p / 10^8 M_\odot)^{1/3}(Nm_*/3\times 10^9
  M_\odot)^{-1/3} (r_c/100\mbox{ pc})$ [$M_p$: 
  primary (more massive) BH mass; $N$: number of stars in the central stellar
  core; $m_*$: average stellar mass; $r_c$: radius of the stellar
  core].

\item The binary orbit is hardened by restricted $3-$body interactions
  with stars during which the stars are ejected at velocities that are
  comparable to the orbital velocity of the binary. The binary becomes
  ``hard'' when its orbital velocity exceeds the velocity dispersion
  $\sigma$ of the ambient stars, literally at $r_h$ $\sim$ $2.7$ pc $(1+q)^{-1}$ $({M_s}/{10^8 M_\odot})$ $({\sigma}/{200 \mbox{ km~s}^{-1}})$ [$M_s$: secondary (less massive) BH mass;
  $q$: mass ratio $M_s / M_p$; \citealt{2005merritt}].  The typical
  hardening radius is $\sim 1$ pc for BH binaries with comparable
  masses of $\sim 10^8 M_\odot$ in each, in a stellar core with a
  typical velocity dispersion of $200$ km~s$^{-1}$.

\item Gravitational wave (GW) radiation becomes dominant in extracting
  angular momentum from the binary when the semi-major axis of the
  binary is less than $r_{GR}$ $\sim$ $0.016 \mbox{ pc }$ $q^{1/4}$
  $(M_p / 10^8 M_\odot)$ $(t_h / 10^8 \mbox{ yr})^{1/4}$ ($t_h$ is the
  hardening timescale in step 2). For BH binaries with masses $M_{\rm BH} \approx 10^8
  M_\odot$ and fiducial values for the hardening timescale $t_h \sim
  10^7$ -- $10^8$ years, this critical radius is around $10^{-3}$ to
  $10^{-2}$ pc.

\end{enumerate}

It is still uncertain whether the transition from (2) to (3) can occur
in less than a Hubble time. The SMBH binary will run out of stars to
eject at a separation of $\sim 1$pc in an axisymmetric stellar
potential \citep{1996quinlan}, so the extraction of angular momentum via
star ejection is no longer efficient for separations of the binary
BHs on sub-parsec scales. This is the so-called ``final parsec
problem'' \citep{1980begelman, 2002Milosavljevic}.  Many mechanisms
have been proposed to overcome the final parsec problem, including gas
accretion onto the binary \citep{2005armitage, 2000gould, 2004escala, 
2008macfadyen, 2009cuadra, 2012rafikov}, refilling the
loss cone by star-star encounters \citep{2005yu, 2005merrittwang}
and/or distorted stellar orbits in triaxial nuclei \citep{2004merritt,
  2011merritt}. While it remains uncertain whether these mechanisms
lead to a final coalescence within a Hubble time, it is certain that
the binary will spend the longest time at sub-parsec separations
\citep{1980begelman}. In principle, sub-parsec SMBH binary systems may
be very common.  Unfortunately, in practice it has proven difficult to
obtain definitive observational evidence of their existence.

\subsection{Observational Evidence of SMBH Pairs/Binaries: Previous Work }

There is indirect evidence for the presence of SMBH pairs at $\sim $
kpc-scales in single galaxies detected via both direct imaging
\citep{2001junkkarinen, 2003komossa, 2004ballo, 2009comerforda,
  2011Fabbiano} and double-peaked narrow emission lines
\citep{2009comerfordb, 2009wang, 2010smith, 2010liub, 2010liua,
  2011shen}. A pc-scale BH pair CSO 0402+379 was detected with the
Very Long Baseline Array \citep{2004maness, 2006rodriguez}, and OJ 287
is believed to be a sub-pc scale SMBH binary due to its periodic
luminosity variations with a period of $\sim 11.86$ yr
\citep{1994takalo, 2000pursimo, 2008valtonen}.  Indirect evidence for
sub-pc--scale pairs includes the precession of radio jets
\citep{1980begelman, 2007liufk}, double-double radio galaxies
\citep{2000shoenmakers, 2003liuwu}, the X-shaped radio sources
\citep{2002dennett, 2002merrittekers, 2004liufk} and ``light
deficits'' in the cores of elliptical galaxies \citep{2002milomerritt,
  2002ravindranath, 2004graham, 2006merritt, 2009kormendy}.

However, there are no confirmed detections of sub-parsec SMBH binaries
to date \citep{2011eracleous}, although they are theoretically
long-lived.  \citet{1996gaskell} interpreted 3C 390.3 with
double-peaked broad emission lines as two SMBHs orbiting around each
other, both associated with a broad line region (BLR). However, long-term
monitoring of this object disproved the binary BH interpretation
\citep{1997eracleous, 2001shapovalova}. In general, Doppler motions in
an accretion disk is favored to explain the double-peaked
broad-emission line quasars \citep{2003eracleous, 2003strateva,
  2004wuliu, 2007gezari, 2010lewis}.

As an alternative means to identify BH binaries via their orbital
motion, there have been a number of searches for large velocity
displacements between single-peaked broad Balmer (mostly H$\beta$)
emission lines and the rest-frame of the quasar host as traced
(typically) by the narrow emission lines. These velocity shifts may
signal orbital motion in a binary. Promising individual candidates
include 3C 227 and Mrk 668 \citep{1983gaskell, 1984gaskell}, SDSS
J092712.65+294344.0 \citep{2009dotti, 2009bogdanovic, 2009vivek,
  2009shields, 2009heckman, 2009decarli}, SDSS J153636.22+044127.0
\citep{2009boroson, 2009chornock, 2010chornock, 2009gaskell,
  2009wrobel, 2010wrobel, 2009decarlidotti}, SDSS J093201.60+031858.7
\citep{2011barrow}, 4C+22.25 \citep{2010decarli}, and SDSS 0956+5128
\citep{2012steinhardt}. For these initial extreme objects, alternate
explanations to BH binaries have been proposed, including recoiling
BHs \citep{2008komossa, 2011eracleous, 2012steinhardt} or a perturbed
accretion disk around a single black hole \citep{2010chornock,
  2010gaskell}.

There have now been two systematic searches through the SDSS for
single-peaked broad emission lines with large velocity offsets from the
quasar rest frame. \citet{Tsalmantza:2011} found five new candidates
with large velocity offsets in Sloan Digital Sky Survey Data
Release 7 (SDSS DR7; \citealt{2000york}) quasars at $0.1< z <
1.5$. \citet{2011eracleous} found 88 candidates with a broad-line peak
offset $\Delta v > 1000$ km s$^{-1}$ out of the 15,900 $z<0.7$ quasars
from SDSS DR7, corresponding to binaries with $r \sim$ few $\times
0.1$ pc. To explain these large velocity offsets in the context of a
binary system, it is necessary to assume that only one of the BHs in
the pair is actively accreting, and that we are catching the pair
close to their maximum separation. It is theoretically plausible that
only one BH in the pair should radiate since binary BHs will torque
the circumbinary disk and open a cavity, analogous to type II planet
migration \citep{1986lin}.  This cavity may at least partially screen
ambient gas from the primary black hole \citep{2008bogdanovic,
  2009cuadra, 2007hayasaki, 2008hayasaki}.
  
After identifying candidate BH binaries from the SDSS spectroscopy
based on large velocity offsets, Eracleous et al.\ (and
\citealt{2013decarli}) have obtained a second epoch of spectroscopy to
search for the expected radial velocity offsets from orbital
motion. They identify 14 candidates with velocity shifts between the
SDSS and second-epoch spectra due to orbital motion of the BHs, whose
resulting accelerations lie between $-$120 and +120 km~s$^{-1}$
yr$^{-1}$. Their methodology is very powerful, but comes with a few
limitations. First, as we mentioned, the requirement of a large
velocity offset ensures that the putative binary system is at its
maximum velocity offset and thus the minimum acceleration in the
orbit.  It is necessary to wait a long time for these systems to
experience significant velocity changes to truly confirm their
nature. Second, the method requires a robust determination of the
systemic velocity of the system: at high redshift, where most of the
quasars and most of the mergers are, narrow emission lines are weak to
nonexistent, and so this technique is limited to $z < 0.8$. Third, and
most important, the method selects extreme objects with highly
nonstandard broad lines.  While these may be the result of binary
motion, they may also be the result of highly uncertain BLR physics.
Fourth, much like double-peaked AGN are seen to vary perhaps due to
hot spots in the disk (e.g., \citealt{2007gezari}), these sources show
changes in their line {\it shapes} when the line luminosity changes,
providing a significant source of noise in the determination of
orbital motion (e.g., \citealt{2013decarli}).  For all of these
reasons, we here pursue a complementary approach with no preselection
on line shape.

\subsection{Our Search Strategy}

Here we propose a complementary approach to that of
\citet{2011eracleous} and \citet{2013decarli}, to search for SMBH
binaries using temporal changes in the velocity of broad emission
lines \citep[see also][]{2013shen}. Specifically, we will search
for radial velocity shifts in normal, single-peaked broad emission
lines in all quasars with multi-epoch spectroscopy. In the past, such
time-resolved spectroscopy was unavailable, leading to
preselection of extreme velocity outliers.  However, with modern
multi-object spectrographs, time-domain surveys are becoming
routine. In the SDSS main survey, $>2000$ quasars with Mg~{\tiny II}
have been observed multiple times, and we focus on these in our pilot
experiment here.  We emphasize that while we begin with this modest
sample, there are already $>8000$ quasars with multiple epochs of
spectroscopy observed with the Baryon Oscillation Spectroscopic Survey
(BOSS, \citealt{2009schlegel}) to which we can apply the same method,
and that upcoming surveys such as an extended BOSS survey and the Prime
Focus Spectrograph on Subaru (PFS, \citealt{2012ellis}) may provide
many more.

Our approach is sensitive to scenarios where either each BH has its
own BLR or only one of the BHs is accreting. In the first scenario,
where each BH has its own BLR, the broad lines are double-peaked. In
this case, the orbital motion of the binary would lead to shifts in
velocity of the double peaks or at least distortion of the broad
emission lines (e.g. from double-peaked to single-peaked lines when
the line-of-sight velocities of the binary BHs decrease from maximum
to zero during 1/4 of their orbit, \citealt{2010shenloeb}). Our
approach is sensitive to these types of line-shape changes in
principle, while search methods looking for persistant double peaks or
extremely high velocity offsets between broad and narrow lines may
miss these systems.

If only one of the binary BHs is accreting, the single-peaked broad
lines will trace the orbital motion of one of the binary BHs. Our
method favors the phases of the binary orbit with large drifts
in the projected radial velocity (i.e., acceleration) of the secondary BH,
while previous studies are tuned to catch the maximum projected
orbital velocity along the line of sight. Since we do not require
narrow lines to identify the velocity rest-frame, we can in principle search AGNs
at all redshifts by studying broad emission lines in the rest-frame
optical and ultraviolet part of the AGN spectra (e.g. Mg~{\tiny II}, C~{\tiny III]},
C~{\tiny IV}, Ly~$\alpha$, etc.).

Of course, our approach has disadvantages and complications as well.
Like \citet{2011eracleous} and \citet{2013decarli}, we are sensitive
to intrinsic variability in the broad emission lines.  Also, because
our targets are so luminous, we are heavily biased towards near-equal
mass ratios, limiting the demographic reach of our survey.  Finally,
and most difficult to quantify, we rely on each BH having its own
broad-line region.  If the BLR surrounds both sources, we will no
longer be sensitive to radial velocity shifts; given the major
uncertainties in BLR physics, this is a major caveat.

Our paper will proceed as follows. We describe the selection and basic
properties of our sample in \S \ref{sec:sample}, and introduce the
cross-correlation method in \S \ref{sec:cross}. The results of our
experiment are presented in \S \ref{sec:candidateselection} and we
estimate the observability of SMBH binary systems with our detection
strategy assuming various scenarios for their evolution in \S
\ref{sec:observability}, and predict the observability with data from
next generation surveys in \S \ref{sec:next}.  We summarize and
conclude in \S \ref{sec:summary}.

%%%%%%%%%%%%%%%%%%%%%%%%%%%%%%%%%%%%%%%%%%%%%%%%%%%%%%%%%%%%%%%%%
%%%%%%%%%%%%%%%% Sample Properties   %%%%%%%%%%%%%%%%%%%%%%%%%%%%%%%
%%%%%%%%%%%%%%%%%%%%%%%%%%%%%%%%%%%%%%%%%%%%%%%%%%%%%%%%%%%%%%%%%
\section{Sample Selection and Properties}
\label{sec:sample}

We will use the quasar catalog from the SDSS to search for binary
black holes.  Specifically, we will focus on quasars with multiple
observations and search for velocity changes in the broad line region
(BLR).  We start with the quasar catalog from DR7
\citep[][]{2010schneider} with 147,330 spectroscopically confirmed
quasars (with no luminosity cutoff), among which 6996 quasars have
multiple observations (quasars are identified with SDSS IDs and
different observations for each quasar are labeled by different
plate-MJD-fiber IDs in SDSS spectra). These objects were targeted
based on their nonstellar colors in $ugriz$ broad-band photometry and
by matching the unresolved sources to the FIRST radio catalogs
\citep{2002richards}.  The calibrated digital spectra from SDSS DR7
have a wavelength range of 3800 - 9200 \AA\ with spectral resolution
of $R \approx 2000$, and each spectral pixel has a width of $\sim 69$
km~s$^{-1}$.  

In some cases, the SDSS has taken multiple spectra of the same
quasars, which gives us the chance to search for orbital motion of the
central BHs. There are several strong broad lines in the rest-frame
optical and UV, including H$\beta$ $\lambda 4861$, Mg~{\tiny II}
$\lambda 2798$ and C~{\tiny IV} $\lambda 1549$. Both the H$\beta$ and
C~{\tiny IV} broad lines have additional sources of complexity when it
comes to measuring their radial velocity. The high-ionization C~{\tiny
  IV} line emerges from the very inner region of the AGN broad line region (BLR)
and often shows asymmetry and absorption that might be
related to disk outflows \citep[e.g.][]{2000proga, 2002richards,
  2005baskin, 2012filiz, 2013capellupo}. For the moment, we will avoid
using C~{\tiny IV}.

The H$\beta$ and Mg~{\tiny II} lines come from further out in the BLR,
where the gas is cooler \citep{2012Goad}. Indeed, in principle
H$\beta$ is attractive because the strong narrow emission from both
H$\beta$ and [O~{\tiny III}] $\lambda \lambda 4959, 5007$ could serve
as a velocity reference \citep{2013shen}. However, we experimented with fitting the
narrow emission lines, and subtracting them to expose the H$\beta$
broad line. We found that the shape of the residual H$\beta$ broad
line can be very uncertain, compromising the subsequent measurement of
velocity shifts. In any case, the number of objects with multiple
observations in the lower redshift range needed to measure H$\beta$ is
small.  Therefore, we focus on the Mg~{\tiny \Rmnum{2}} broad lines as
our diagnostic of orbital motion of the secondary black hole in this
work.

To keep Mg~{\tiny \Rmnum{2}} in the wavelength window, we select
quasars with $0.36<z<2.0$ and multiple observations. While $2.0<z<2.3$
quasars also have the centroid of Mg~{\tiny \Rmnum{2}} in the 3800 -
9200 \AA\ window, the data near the red end of the SDSS spectra suffer
from sky subtraction errors, which are severe contaminants.  We thus
exclude the $2.0<z<2.3$ quasars from our analysis. We also reject
spectra with more than 100 pixels of missing data (as flagged with a
zero in their inverse variance spectra). Finally we have a sample of
4204 quasars. We have not explicitly rejected spectra from
  `marginal' or `bad' plates.  These flags indicate that the S/N may
  be low on the plate or data may be missing from many of the
  spectra. We pay extra attention to those candidates with marginal or 
  bad quality flags. Indeed, many of the plates in our sample may have been
  reobserved due to `marginal' quality originally. Finally, as we will state in Section
\ref{sec:candidateselection}, we find that spectra with low signal to
noise ratio (S/N) confuse our cross-correlation procedure, so we
construct a high S/N sample of 1523 objects with S/N pixel$^{-1}$ $>$
10 to focus on the most reliable cross-correlation measurements. Our
final candidates will be selected from the high S/N subsample of 1523
objects.

\begin{figure*}[ht]
	\centering
	\includegraphics[width=0.45\textwidth]{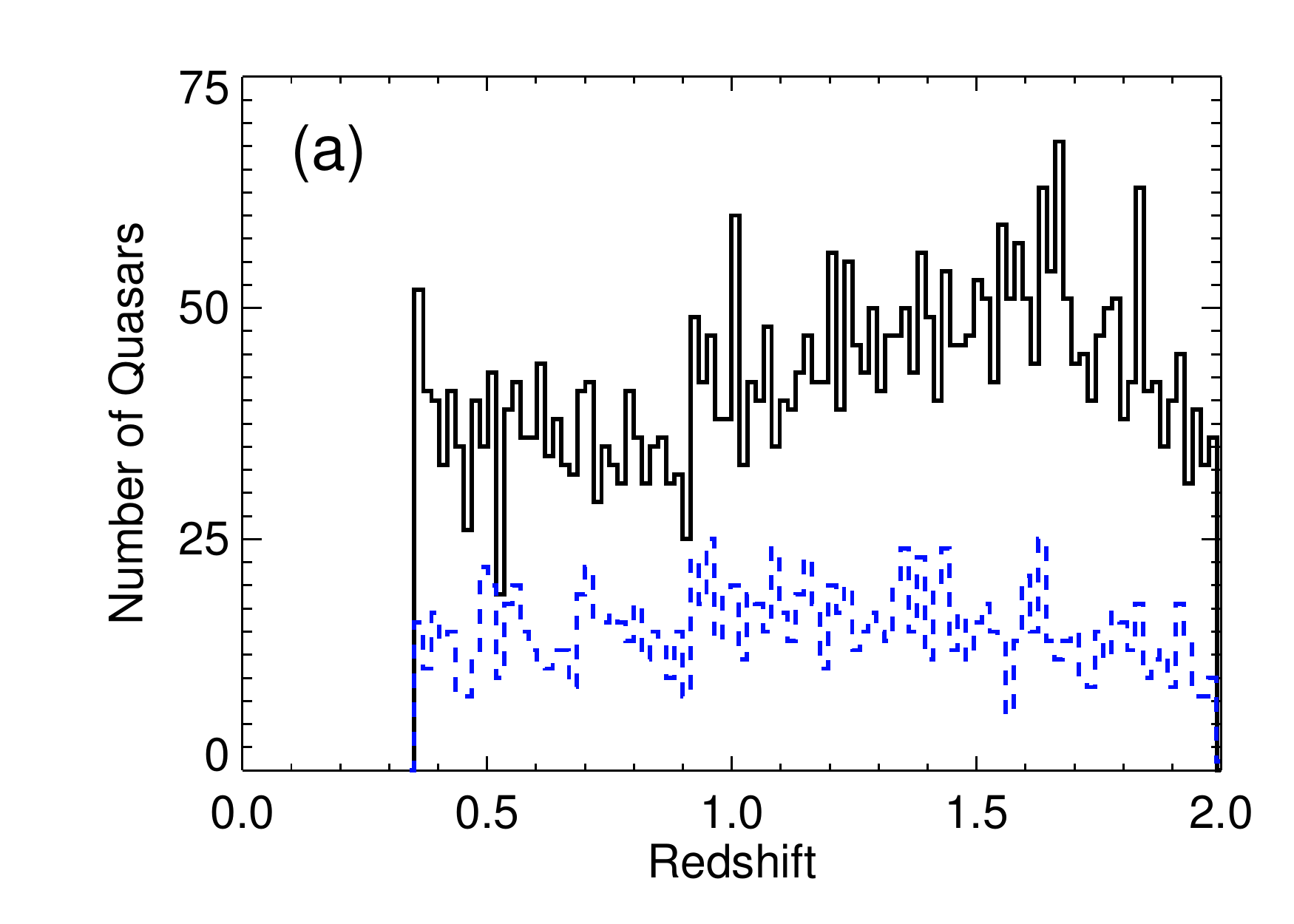}
	\includegraphics[width=0.45\textwidth]{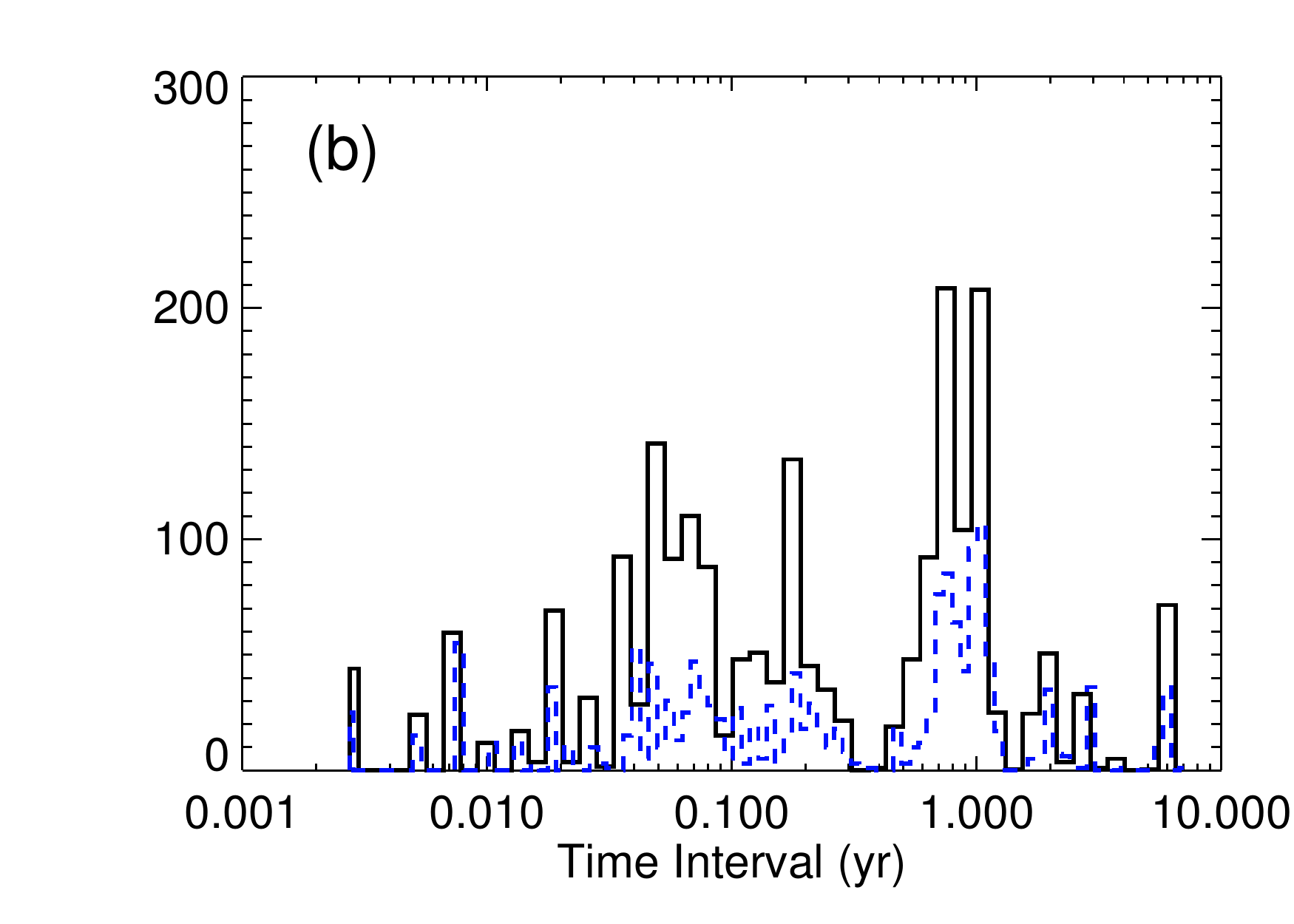}
	\includegraphics[width=0.45\textwidth]{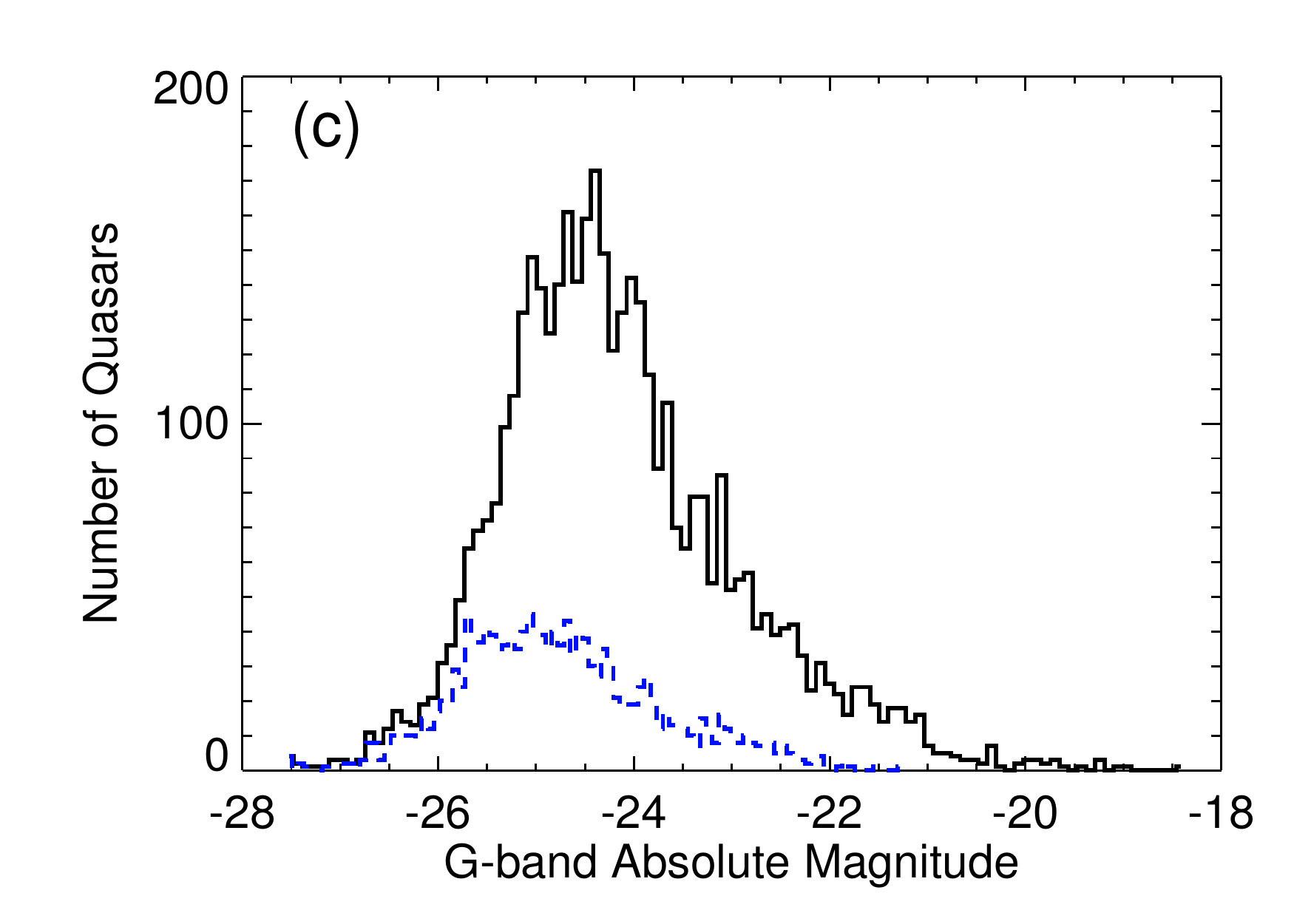}
	\includegraphics[width=0.45\textwidth]{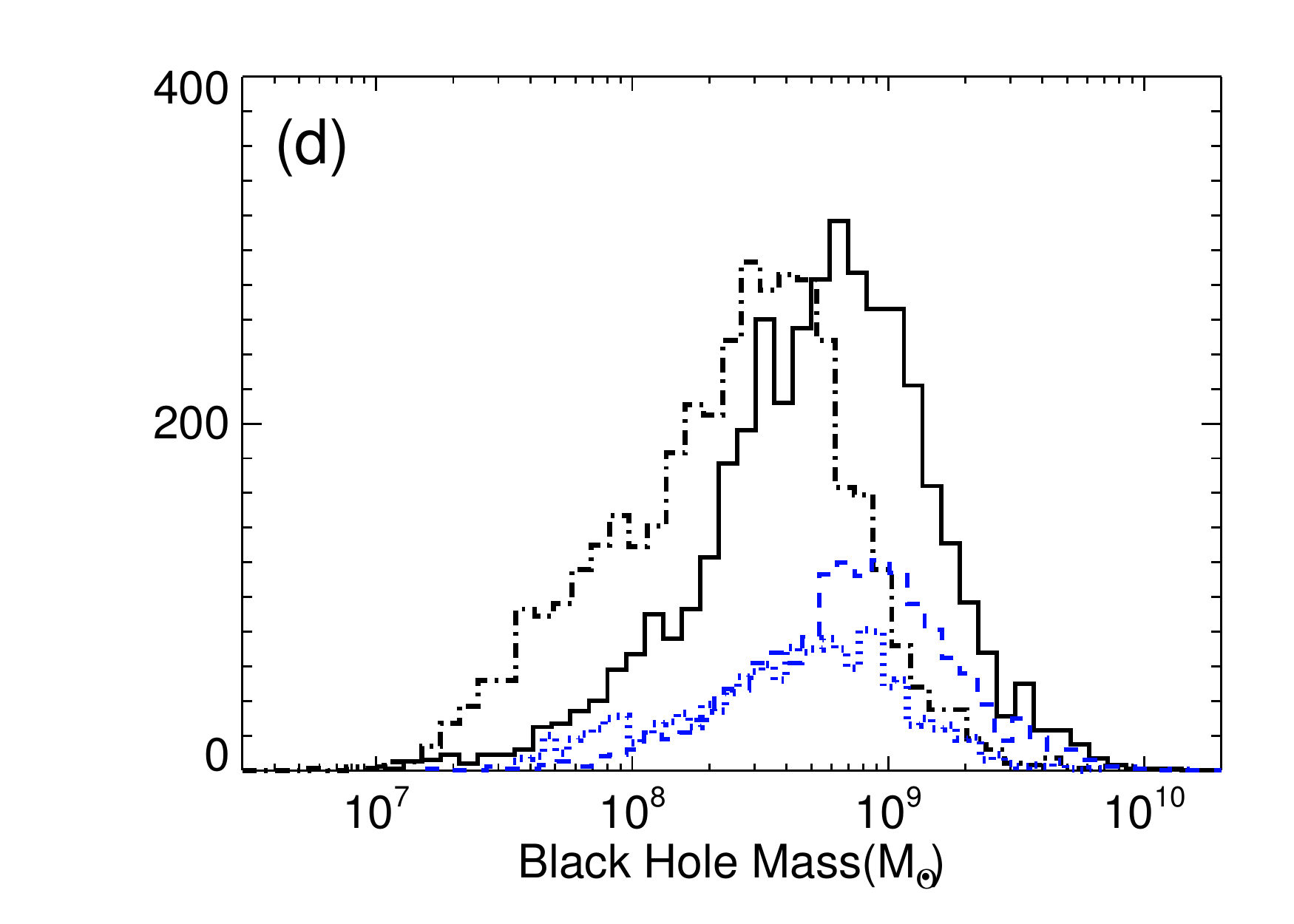}
\caption{Basic properties of our sample: (a) Redshift. We select the
 redshift range of [0.36, 2] to keep the Mg~{\tiny II} line in the
 observing window of SDSS. (b) Time interval between multiple
 observations of the quasars. (c) Absolute magnitudes. (d) BH mass
 distribution. 
 For (a), (b) and (c), the black (solid) lines are for the whole
 sample while the blue (dashed) lines are for the S/N per pixel $>$ 10
 subsample. For (d), the black (solid and dash-dotted) lines show the
 distribution of BH masses for the whole sample while the blue (dashed
 and dotted) lines are for the high S/N subsample. The solid and
 dashed lines are based on Mg~{\tiny \Rmnum{2}} widths
 \citep{2011shen2} while the dash-dotted and dotted lines show lower
 limits on the BH masses derived from their bolometric luminosities
 \citep{2011shen2} assuming that the bolometric luminosities are 50\%
 of their Eddington luminosities. We assume that these BH masses are the secondary 
masses in a BH binary in our working scenario.}
\label{fig:SampleInfo}
\end{figure*}

The basic properties of the whole sample and the high S/N subsample
are shown in Fig. \ref{fig:SampleInfo}. The redshifts
(Fig. \ref{fig:SampleInfo}a) are measured by template fitting of the
quasar spectra with least-squares linear combinations of redshifted
template eigenspectra \citep{2012bolton}. The mean redshift is
$\langle z \rangle = 1.22$. The mean time interval
(Fig. \ref{fig:SampleInfo}b) between two observations of the same
quasar is 0.7 years with a range from 0.003 to 7.03 years. The
$g$-band absolute magnitude of the sample (Fig. \ref{fig:SampleInfo}c)
ranges from $-18.4$ to $-27.5$ mag with a median value of $-24.1$ mag. The
blue dashed lines in Fig. \ref{fig:SampleInfo}a, \ref{fig:SampleInfo}b
and \ref{fig:SampleInfo}c are the corresponding property distributions
of the high S/N subsample. 

In Fig. \ref{fig:SampleInfo}(d), the black (solid and dash-dotted)
lines are BH mass distributions for the 4047 quasars in the whole
sample (157 AGNs don't have available BH masses; they are assigned the
median BH mass), while the blue (dotted and dashed) lines are for the
high S/N subsample (6 of which do not have BH
masses). We present two estimates of the BH mass.  The solid and the
dashed lines represent BH masses measured from the Mg~{\tiny
  \Rmnum{2}} broad-line widths from \citet{2011shen2}.  They range
from $10^{7}$ $M_\odot$ to $10^{10.5}$ $M_\odot$ and are peaked at
$10^{8.65}$ $M_\odot$ (or $10^{8.9}$ $M_\odot$) for the whole sample
(high S/N subsample). Taking the BH masses at face value together with
the bolometric luminosities of these quasars \citep{2011shen2}, the
Eddington ratios ($\dot{m}_{\rm Edd}$) have a median value of $\sim
0.25$ while ranging from 0.02 to 2.7.  The dash-dotted and dotted
lines are lower limits on the BH mass derived from the bolometric
luminosities \citep{2011shen2}, assuming that the bolometric
luminosity of each quasar is $50\%$ of its Eddington luminosity. They
range from $10^{6.3}$ $M_\odot$ to $10^{9.8}$ $M_\odot$ and are peaked
at $10^{8.3}$ $M_\odot$ (or $10^{8.7}$ $M_\odot$) for the whole sample
(high S/N subsample).

%%f1

%%%%%%%%%%%%%%%%%%%%%%%%%%%%%%%%%%%%%%%%%%%%%%%%%%%%%%%%%%%%%%%%
%%%%%%%%%%%%%%%%%%% Cross-Correlation %%%%%%%%%%%%%%%%%%%%%%%%%%%%%%%
%%%%%%%%%%%%%%%%%%%%%%%%%%%%%%%%%%%%%%%%%%%%%%%%%%%%%%%%%%%%%%%%
\section{Cross-Correlation Analysis}
\label{sec:cross}

%%%%f2
\begin{figure*}[ht]
	\includegraphics[width= 0.45\textwidth]{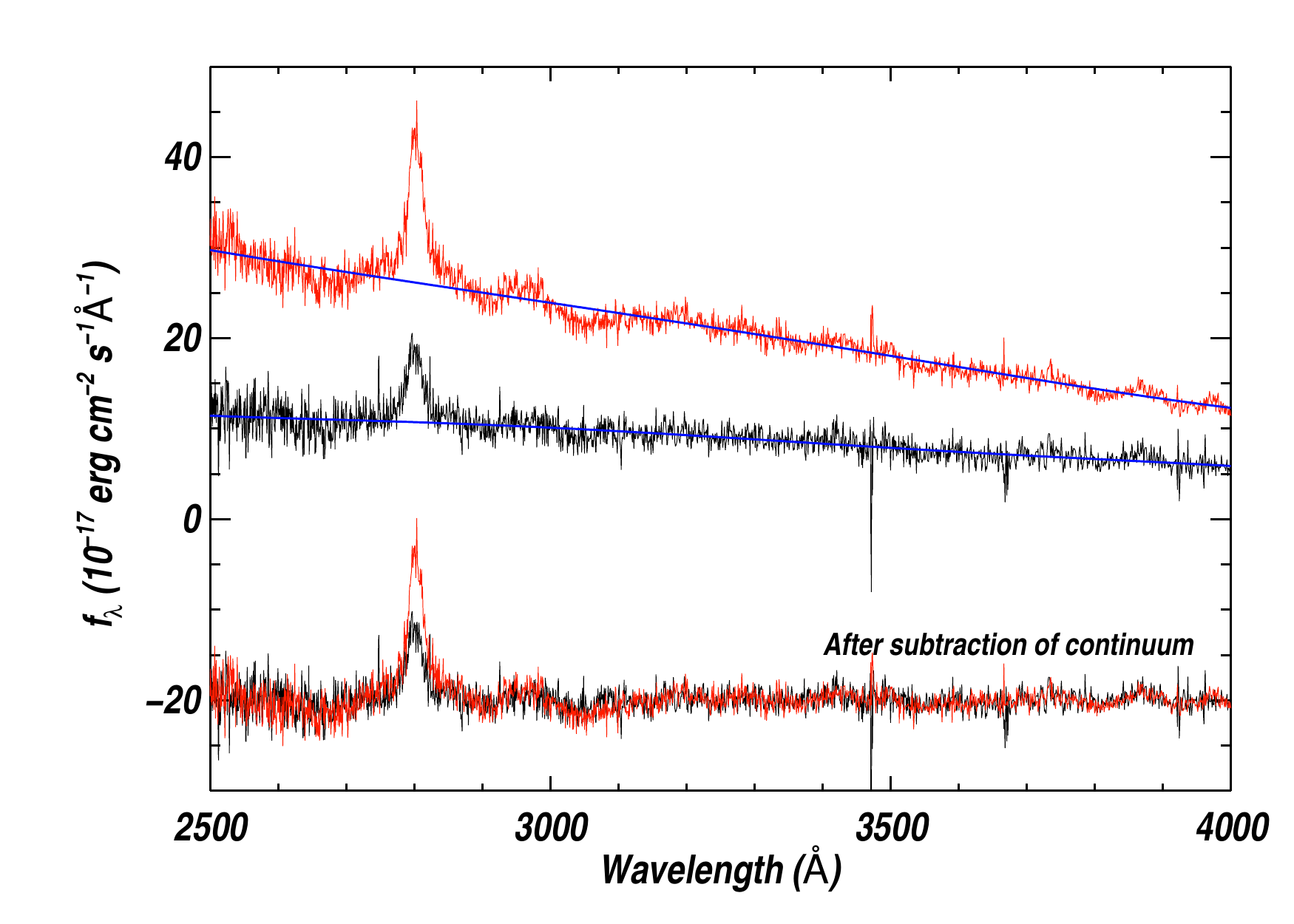}
	\includegraphics[width= 0.45\textwidth]{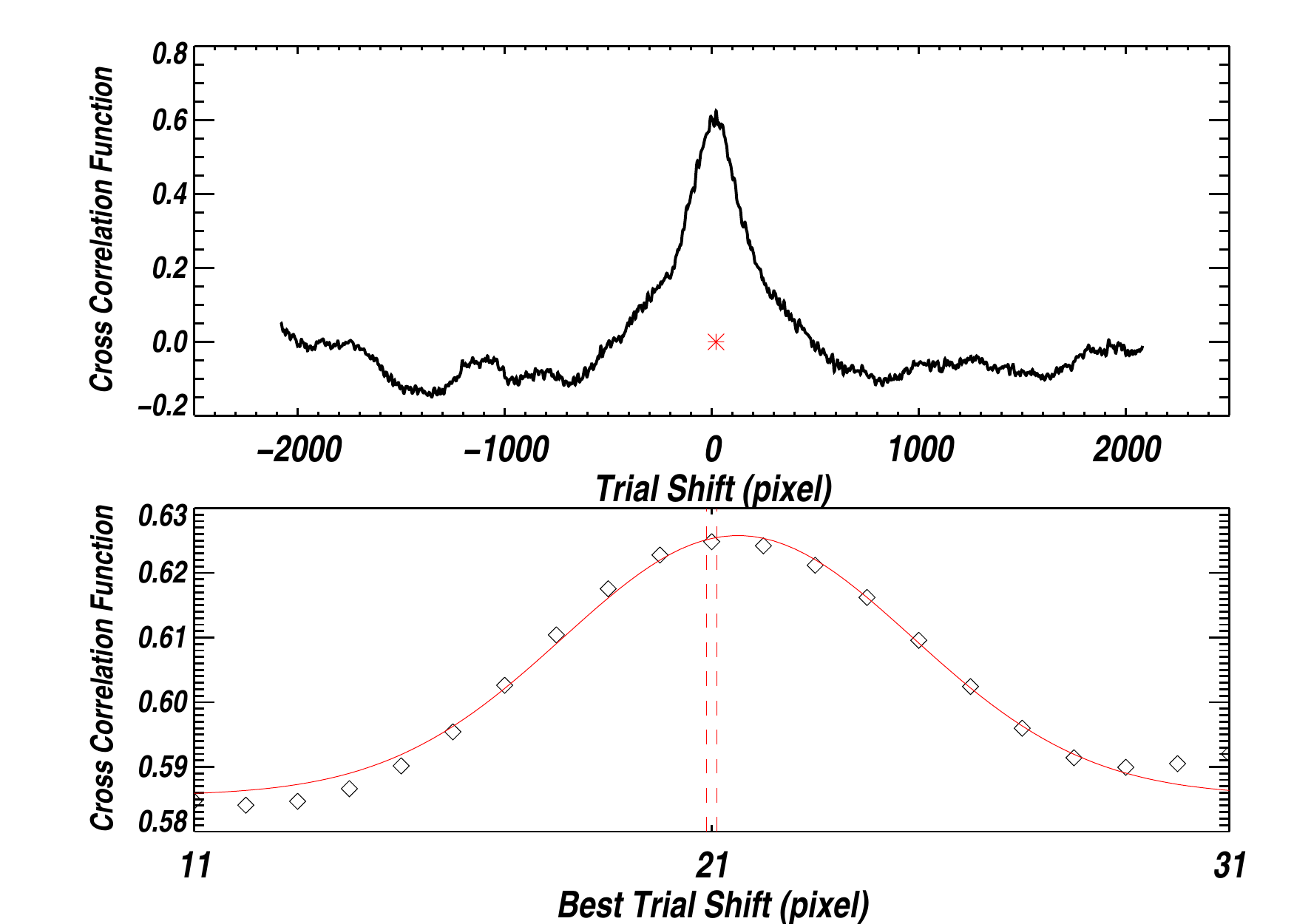}
\caption{
  Example of the analysis of SDSS J095656.42+535023.2: (a) Polynomial
  fitting and subtraction of the continuum. The black and red lines on
  the upper part of the figure show the spectra of SDSS
  J095656.42+535023.2 at two epochs with a 6.16 year separation. The
  blue lines are the fitted continuum. Spectra after continuum
  subtraction are shown on the bottom of the figure. (b)
  Cross-correlation function and peak hunting. The upper panel shows
  the cross-correlation at different trial shifts in pixels. The lower
  panel is zoomed in on 20 pixels at the peak region of the
  cross-correlation function. The two vertical dashed lines mark the
  position of the peak. The width of 1.7 km~s$^{-1}$ is the
  uncertainty in the peak reported by the Gaussian fitting. The
  resolution, i.e. the interval between two neighboring pixels (open
  square dots), is 17.3 km~s$^{-1}$. The best trial shift is marked
  with a red star on the upper panel.  }
\label{fig:example}
\end{figure*}

We are searching for a change in velocity in the Mg~{\tiny II} line
linked to orbital motion in a sub-pc BH binary.  Note that the SDSS
measures a redshift for every quasar at every epoch as well. In
principle, we could use these redshift measurements to find cases of
large radial velocity shifts. However, the SDSS redshift measurements
have a larger dispersion from one epoch to the next than our
method. Therefore, we use a single redshift for each quasar to move
the spectra to the rest frame. All the following analyses are
performed in the rest frame of the quasar.

In order to isolate this signal, we must isolate the broad emission
line from the continuum of the quasar. We find that variability in the
quasar continuum can significantly decrease the sensitivity of the
cross-correlation to the broad-line velocity shift. We fit the
continuum with a fifth-order polynomial. We prefer this to a power-law
because in some spectra the blue end of the continuum drops
significantly due to an artifact, making it impossible to fit with a
single power-law. In fitting the continuum, we mask all the main
emission lines and the $2\sigma$ wavelength regions around the
centroids in the range from 1000 \AA\ to 7300\AA\ (the centroid
wavelengths and widths are from \citealt{2001vandenberd}, Table 2),
namely: O {\tiny VI} (1000 \AA\ -- 1070 \AA), Ly $\alpha$ (1142 \AA\
-- 1290 \AA), Si {\tiny IV} (1389 \AA\ -- 1411 \AA), C~{\tiny IV} (1526
\AA\ -- 1572 \AA), C {\tiny III}] (1892 \AA\ -- 1926 \AA), Mg~{\tiny
  II} (2770 \AA\ -- 2826 \AA), H$\gamma$ (4318 \AA\ -- 4364 \AA),
H$\beta$ (4810 \AA\ -- 4912 \AA), [O {\tiny III}] (4928 \AA\ -- 5038
\AA), He I (5800 \AA\ -- 6000 \AA, 7000 \AA\ -- 7200 \AA) and
H$\alpha$ (6300 \AA\ -- 6800 \AA). We subtract the fitted continuum to
expose the {\mgii} broad lines. One example is shown in Figure
\ref{fig:example}a.

Secondly, we cross-correlate the continuum-subtracted spectra from the
two epochs, using all data in the wavelength range $\lambda =$ 1700
\AA\ -- 4000 \AA. We weight the flux in each pixel with its inverse
variance from the SDSS observation, which efficiently prevents bad
pixels from confusing our analysis. We calculate the cross-correlation
at various trial velocity shifts:
\begin{equation}
P_{cross-corr}(D) = \frac{\sum_{k=0}^{N-|D|-1} (a_{k+|D|} - \bar{a})(b_k - \bar{b})}{\sqrt{ [\sum_{k=0}^{N-1} (a_k - \bar{a})^2] [\sum_{k=0}^{N-1} (b_k - \bar{b})^2]}}
\end{equation}
where a and b are the spectra from the two epochs, and D is the trial
delay. The spacing of the trial delays is 17.3 km~s$^{-1}$, $1/4$ of the SDSS
spectral resolution.

Thirdly, we look for the peak in the cross-correlation function and
define the corresponding trial velocity shift as our "best"
measurement of velocity shift. The method we use to find the peak in
the cross-correlation function is adapted from the SDSS III (BOSS)
pipeline \citep{2012bolton}, which measures centroids of peaks using a
Gaussian fit around the minimum $\chi^2$. The peak cross-correlation
values we find are equivalent to the maximum value in most cases, but
in some extreme cases where there are offset local fluctuations
sitting on the "true" peak, this Gaussian fitting method behaves
better. In general, the uncertainty returned from the peak fitting is
small ($< 20$~\kms), and thus much smaller than our systematic
uncertainties (\S \ref{subsec:uncertainty}).  An example of our
cross-correlation and peak hunting is shown in Fig
\ref{fig:example}b.

We mainly focus on the spectral region directly surrounding Mg~{\tiny
  II} (2700 \AA\ -- 2900 \AA), where the strongest
cross-correlation signal can be found.  However, we are throwing away
real information in the weaker broad emission lines such as Fe{\tiny
  II}, which are found throughout the rest-frame near-UV spectrum.
Thus, we also perform a wider fit to the region between 1700 \AA\ and
4000 \AA.  (\S \ref{subsec:uncertainty}). Using two wavelength regions allows us to
remove many false positives that come from the narrow Mg~{\tiny II}
considered alone.

%%%%%%%%%%%%%%%%%%%%%%%%%%%%%%%%%%%%%%%%%%%%%%%%%%%%%%%%%%%%%%%%%
%%%%%%%%%%%%%%%% Candidate Selection %%%%%%%%%%%%%%%%%%%%%%%%%%%%%%%
%%%%%%%%%%%%%%%%%%%%%%%%%%%%%%%%%%%%%%%%%%%%%%%%%%%%%%%%%%%%%%%%%
\section{Results: SMBH Binary Candidate Selection}
\label{sec:candidateselection}

\begin{figure*}[!ht]
	\centering
	\includegraphics[width= 0.47\textwidth]{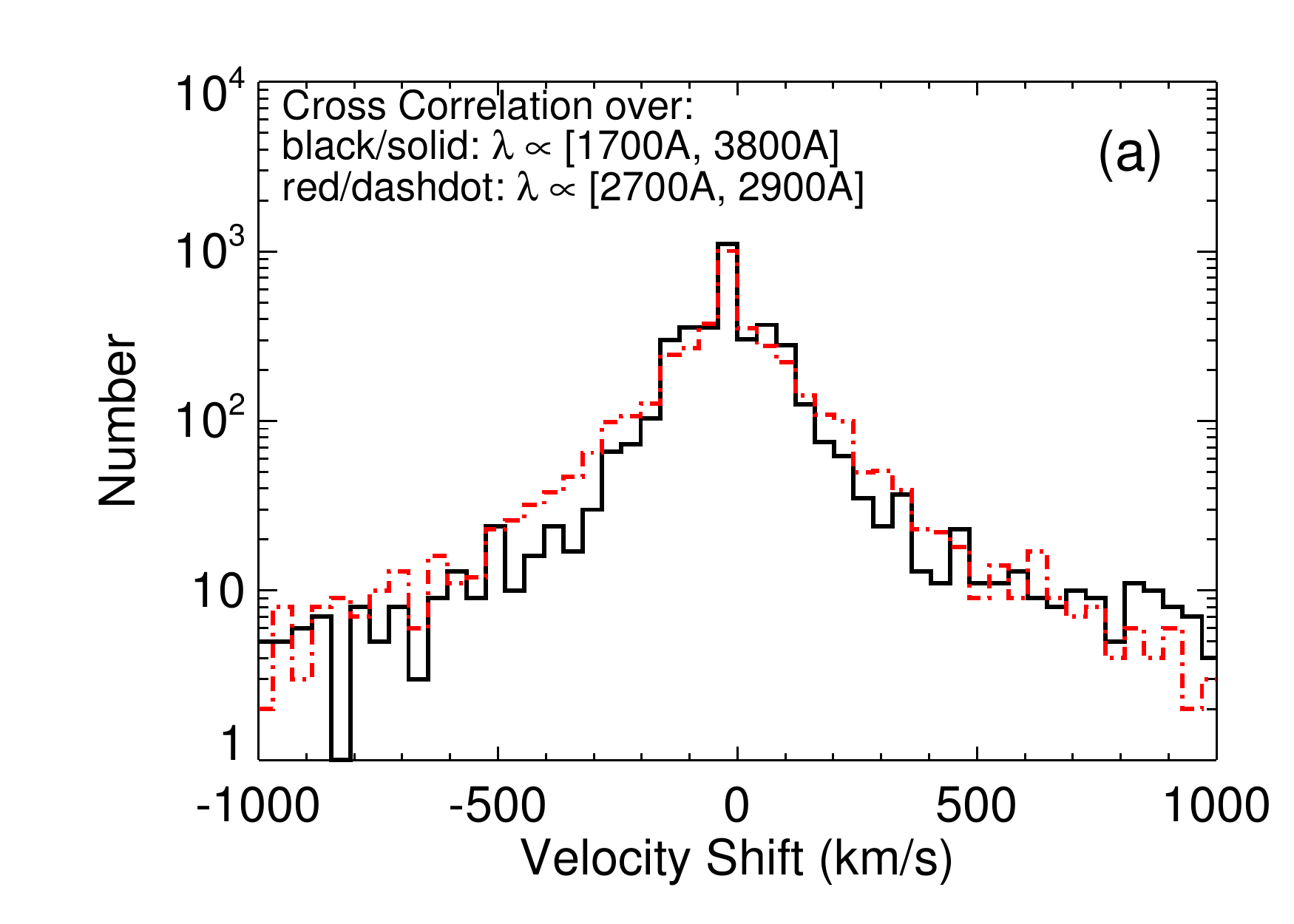}
	\includegraphics[width= 0.47\textwidth]{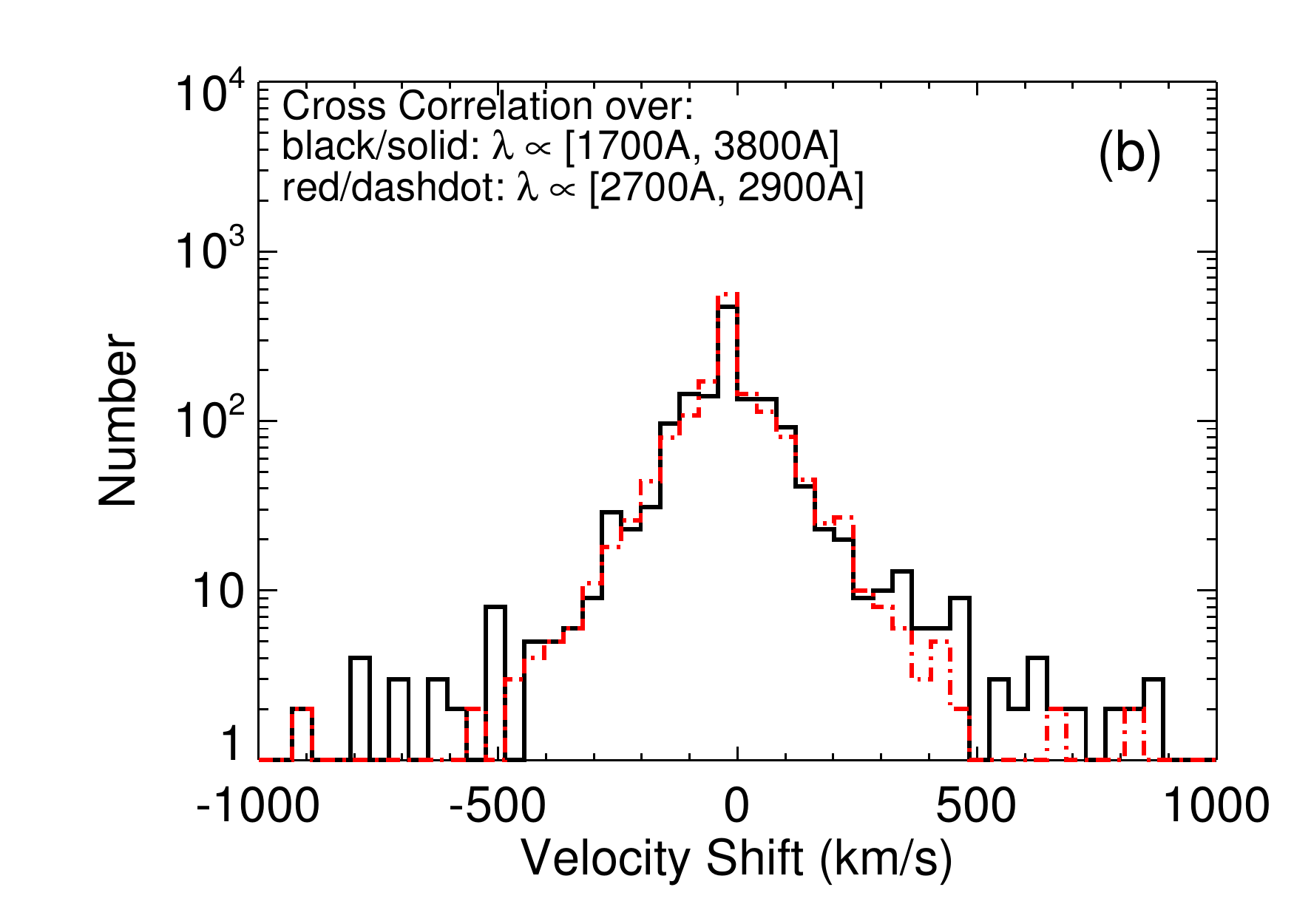}
\caption{ Measured velocity change of the broad line region for: (a)
  the whole sample. The width of the distribution is $\sim$ 101 km~s$^{-1}$
  with cross-correlation over the whole wavelength range [1700 \AA,
  4000 \AA], and is $\sim$ 126 km~s$^{-1}$ with cross-correlation over the
  wavelength range near Mg~{\tiny II} [2700 \AA, 2900 \AA]; (b) the S/N per pixel $> 10$
  subsample. The width of the distribution is $\sim$ 82 km~s$^{-1}$ with
  cross-correlation over both the [1700 \AA, 4000 \AA] and the [2700 \AA, 2900
  \AA] ranges.}
\label{fig: histo-ErrBar}
\end{figure*}

We measured broad-line velocity shifts for all of the 4204 quasars in
our sample, and their distribution is shown in Fig \ref{fig:
  histo-ErrBar}. Since most of the quasars have very short time
intervals between observations, their velocities will not change even
if they harbor binary BHs.  Therefore, the width of this histogram
represents the typical errors incurred from noise in the data, sky
subtraction and wavelength calibration errors, the finite width of the broad lines and intrinsic
variability in broad emission or absorption line shape.  It is peaked
at zero and its width is $\sigma = 101$~km~s$^{-1}$.  In practice, we
find that spectra with low S/N confuse our cross-correlation
procedure, so we construct a high S/N sample of 1523 objects with S/N
pixel$^{-1}$ $>10$ to focus on the most reliable cross-correlation
measurements.  The distribution of measured velocity shifts in the high S/N
sample is shown in Fig \ref{fig: histo-ErrBar}. We
get a histogram width of $\sigma \approx 82$ km~s$^{-1}$ for this
subsample, and accordingly we set $3.4 \sigma \sim 280$ km~s$^{-1}$ as
the lower limit for ``significant velocity drifts'' such that we
expect $<1$ false positive detection in the high S/N sample.  However,
we also insist that the cross-correlation measured from the Mg~{\small
  II} region alone be consistent within $1~\sigma$ to that measured
over the entire spectral region.  By comparing the cross-correlations
of the two regions, we are able to remove apparent shifts that are due
to some small errors in sky subtraction, for instance.

%% f3

In the end, we pick our final list of candidates from the high S/N sample with
the following criteria:

\begin{em}
  Velocity shifts measured with the two methods (cross-correlating the
  whole spectral range or only the Mg {\tiny II} line) should satisfy:
  (1) at least one of the two measurements is larger than 280
  km~s$^{-1}$; (2) the difference between them is less than $1~\sigma$ or 82
  km~s$^{-1}$; (3) the error in the cross-correlation fit is $< 100$~\kms.
\end{em}

Seven objects obey these criteria. The selected candidates are
described in Table \ref{table: selected}. One of our final candidates 
(SDSS J095656.42+535023.2) comes from a 
plate flagged as `bad'. This flag reflects the very poor seeing conditions and bright sky 
when the data were taken, and may lead to large spectrophotometric uncertainty.  However, most 
of the spurious features are introduced in the red end of the spectrum, far from the Mg~{\tiny II} 
line.  We thus believe that our polynomial continuum fit will account for any remaining flux 
calibration issues in this target. Noting that most of the
sources have velocity shifts very close to our limit, we consider
seven as an upper limit to the total number of detections in our
sample. Interestingly, most of our candidates have $z < 1$.  We
suspect that this is due to the increasing difficulties of working in
the red.  The distribution of BH masses is quite similar to that of
the whole sample.  According to these BH masses and the measured
accelerations, the candidate SMBH binaries are expected to have
separations of $0.01 - 0.1$ pc at maximum. The numbers are small, but
we find no abnormal spectral properties among these seven candidates.
For instance, their SDSS colors (e.g., $g - i$) do not seem atypical
for QSOs in this redshift range, as may be expected for a circumbinary
disk \citep[e.g.][]{1995syer, 2012gultekin, 2012rafikov}.

\begin{figure*}[ht]
	\centering
	\includegraphics[width= \textwidth]{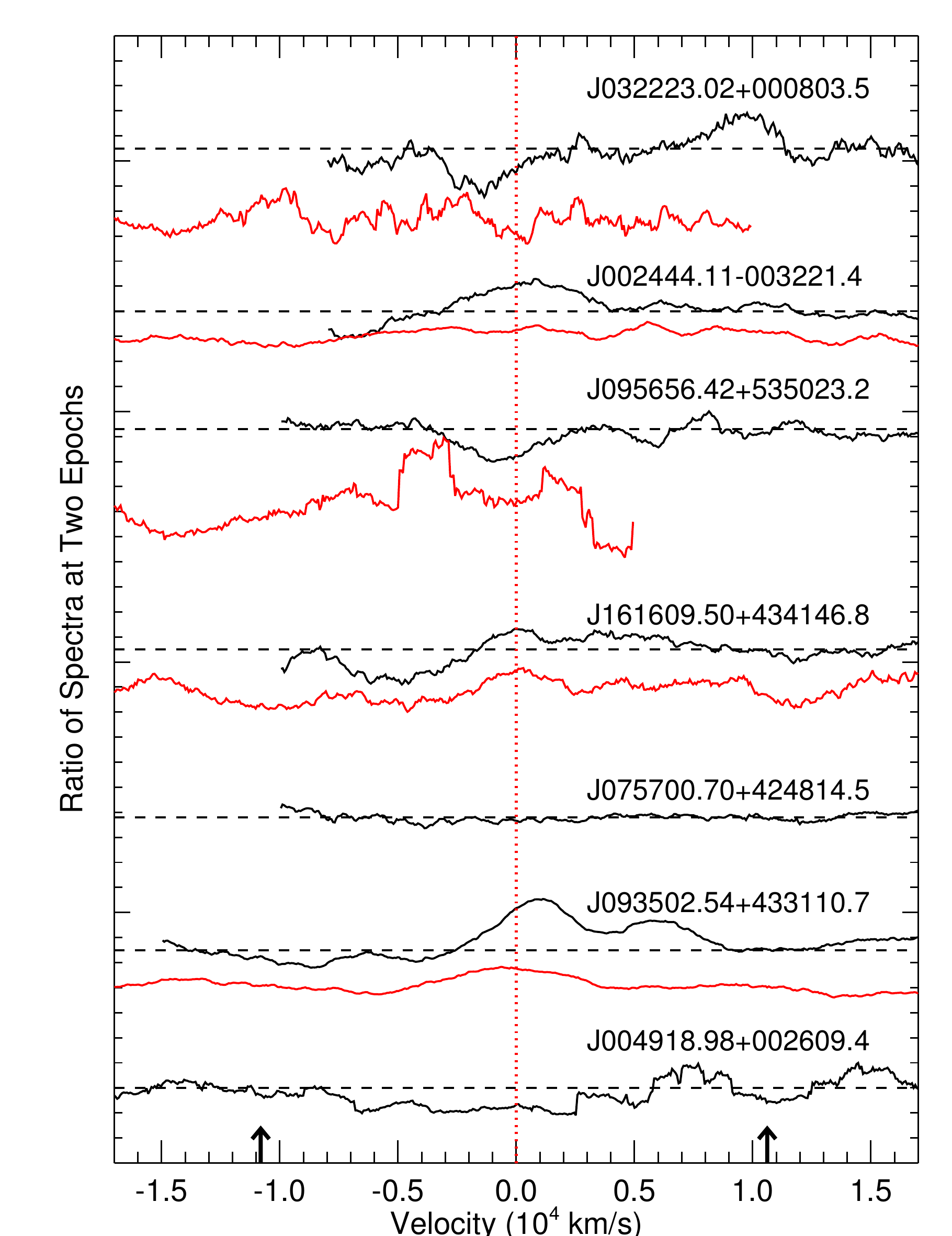}
\caption{
  Ratio of spectra at two epochs of the candidate SMBH binaries. The
  black solid lines show ratios of {\mgii} broad lines, while the red
  solid lines show ratios of H$\beta$ lines for comparison. Zero
  velocity lies at the rest frame of {\mgii} and H$\beta$ lines. The
  horizontal dashed lines show the approximate average levels of the
  ratio spectra far from the emission lines. Five of the seven candidates show ``bumps" and
  ``valleys" around the line peaks in both {\mgii} and H$\beta$,
  consistently indicating the presence of velocity shifts in these
  quasars. The two arrows on the x-axis mark the velocity range where
  we perform cross correlation of the {\mgii} lines.}
\label{fig:ratio}
\end{figure*}

Since the velocity shifts in the {\mgii} broad lines are small
compared with the linewidths, in Fig. \ref{fig:ratio} we plot the
ratio of the spectra at the two epochs for each candidate, at
wavelengths around {\mgii} and H$\beta$ (if present). These ratio
spectra nicely highlight the velocity shifts via ``bumps" or
``valleys" around the line peaks.  We clearly see such features in the
ratio spectra of SDSS J032223.02+000803.5, SDSS J002444.11-003221.4,
SDSS J095656.42+535023.2, SDSS J161609.50+434146.8 and SDSS
J093502.54+433110.7.  One target, SDSS J093502.54+433110.7, is in 
common with \citet{2013shen}.

While using the cross-correlation of two spectral regions effectively
removes spurious targets, it also may remove some real targets.  Thus,
for completeness, in Table \ref{table: selected2}, we list all 64
targets that are recovered with velocity shifts larger than
$280$~\kms\ by the Mg {\tiny II} cross-correlation alone, for future
more detailed follow-up.  Note that none of these targets comes from a 
`bad' plate although a few are from marginal plates as indicated. However, given the many uncertainties
remaining in selecting these targets, we will consider seven objects as
an upper limit on the true number of detected binary systems in our
sample for the purposes of discussion.

%% f4

%%%%%% Uncertainty
\subsection{Uncertainties}
\label{subsec:uncertainty}

There are various sources of uncertainties in our measurements,
including wavelength calibration, sky subtraction, continuum
subtraction and cross-correlation errors. The statistical error
reported by the cross-correlation procedure is typically $<20$~\kms,
with only $\sim 3\%$ of objects having uncertainties greater than
$100$~\kms.  We further create mock spectra using the SDSS variance
arrays, and perform input-output tests of our cross-correlation analysis. 
We recover a very narrow distribution of uncertainties in our
output velocity shifts ($< 34$~\kms\ in all cases).  Random errors alone 
cannot explain the width of the velocity distribution that we observe.

One source of systematic uncertainty comes from our choice of
wavelength region for the cross-correlation. While the cross
correlation over the whole range [1700\AA, 4000\AA] helps us remove
some false-positives, it also introduces further potential sources 
of uncertainty.  For instance, absorption from the Ca
H+K lines around the 4000\AA\ break, or errors in sky subtraction, may
confuse our technique. As a sanity check, we did another set of cross
correlations for the seven candidates over [1700\AA, 3800\AA] to
exclude Ca H+K absorption lines. Only three of the seven candidates
(SDSS J002444.11-003221.4, SDSS J095656.42+535023.2, SDSS
J004918.98+002609.4) survive this test.  These are our three most
reliable candidates.  As a result, we consider seven as an upper limit in
what follows.

There are many other contributions to the velocity uncertainties.  The
first effect is uncertainties in sky subtraction, as already
discussed.  Second, intrinsic changes in the quasar continuum may be a
source of uncertainty. When we change the spectral
region used for the continuum fit, the output velocities can change by
tens of \kms, showing that our continuum fitting also contributes to
the error budget.  Third, there is irreducible scatter from temporal
variations in the broad lines themselves \citep[e.g.,][]{2007sergeev,
  2007gezari,2013decarli}. We do not yet know the full range of velocity 
shifts resulting from line-shape variability.  Therefore, we cannot take the 
observed distribution of velocity offsets and infer anything about the binary 
population below our detection threshold, as in \citet{2012badenes}.

%%%%%%%%%%%%%%%%%%%%%%%%%%%%%%%%%%%%%%%%%%%%%%%%%%%%%%%%%%%%%%%%%
%%%%%%%%%%%%%%%% Observability Estimation %%%%%%%%%%%%%%%%%%%%%%%%%%%%%%%
%%%%%%%%%%%%%%%%%%%%%%%%%%%%%%%%%%%%%%%%%%%%%%%%%%%%%%%%%%%%%%%%%
\section{Expected Number of Observable Candidates}
\label{sec:observability}

We define the {\it observability} of an SMBH binary system to be the
probability for the binary to have a projected velocity drift
above our detection threshold as defined in \S \ref{sec:candidateselection}. In
this section, we estimate the observability of BH binaries given the 
distribution of BH masses and time-intervals in sample, and two different assumptions
about the BH merging process. For simplicity, we start with the assumption that all
the quasars are in a tight binary phase, with only
one accreting BH. We further assume that the quasar activity that we
observe is associated with the secondary BH.  Simulations do find
increased accretion rates onto the secondary from a circumbinary disk
due to the proximity of the secondary to the ambient gas
\citep{2009cuadra}.  Therefore, we assume the BH masses measured by
\citet{2011shen2} shown in Fig. \ref{fig:SampleInfo}d are the masses
of the secondary BHs. Note that since the quasars are quite luminous, they are most 
likely found in comparable-mass binaries under these assumptions.

The observability of SMBH binaries depends on the following
parameters.  (1) The total BH mass [$M_{\rm BH}$, $(1+q) / q$ $\times$
  measured BH mass], mass ratio ($q$ $=$ $M_{\rm secondary}$
  $/M_{\rm primary}$) and separation of the two black holes ($r$), which
  together determine the acceleration of the secondary. (2) The time
interval ($\Delta t$) between multiple observations of the same
quasar. (3) The projection of the orbit on the sky. For our sample, we
know the time interval between observations, and we assume that the
total BH mass of the binary is known.  We assume that the binary
inclination and phase are oriented randomly on the sky with respect to
the observer.  The only free parameters are the mass ratio $q$ and
separation $r$ of the binary black holes.

This section will proceed as follows. In \S \ref{sec:fixaq} we calculate the
observability of a BH binary with a given mass, mass ratio,
separation, observing time interval and orientation. Then we randomly
sample the inclination as well as the phase of the binary, and
calculate the observability of each quasar in our sample utilizing the
BH mass and observing time interval. The sum of observabilities for
all of the quasars gives us the total expected number of observable
candidates ($N_{\rm obs}$). We will derive the dependence of $N_{\rm obs}$ on
mass ratio by fixing the orbital separation and on separation by
fixing the mass ratio. 

We then relax the assumption that the BH binaries spend their whole
life at a fixed separation. In \S \ref{sec:timeevolution}, we
theoretically calculate the residence time ($t_{\rm res}$) as a function
of radius of the BH binaries in the scenario of gas-assisted evolution
at sub-pc scales ($r \lesssim 0.1$ pc). Since the probability to find a
quasar at a given radius is proportional to the residence time of the
BH binary at that radius [$dt = (t_{\rm res} / r) dr$], the detectability
of one SMBH binary system is then the integral of the detectability
over radius, weighted by the residence time:

\begin{eqnarray}
&&P_{obs}(M_{\rm BH}, q, \Delta t, \dot{m}_{\rm Edd}) \nonumber\\
&=&\int_{\rblr}^{0.1 \mathrm{pc}}P_{obs} (r, M_{\rm BH}, q, \Delta t) \frac{t_{\rm res}(r, M_{\rm BH}, \dot{m}_{\rm Edd})}{r} dr,\nonumber\\
&&
\label{eq:rblr}
\end{eqnarray}
where $\Delta t$ is the time interval between observations of one AGN, 
$\rblr$ is the size of the BLR \citep{2010shenloeb}:
\begin{equation}
\rblr \approx 2.2 \times 10^{-2} \left(\frac{\dot{m}_{\rm Edd}}{0.1} \right)^{1/2} \left(\frac{M_{\rm BH}}{10^8 M_\odot}\right)^{1/2} \text{pc},
\label{eq:BLR}
\end{equation}
and $\dot{m}_{\rm Edd}$ is the Eddington accretion rate ratio.  The
$\rblr$ scaling is derived from local reverberation mapping campaigns
\citep[e.g.,][]{2009bentz, 2013bentz}, which find that $\rblr \propto
L^{1/2}$. Note that we are extrapolating this relation into the
redshift and luminosity range of our observed quasars.  The probability
$P_{obs}(M_{\rm BH}, q, \Delta t)$ is evaluated for each object in our
sample using its own $M_{\rm BH}$ and $\Delta t$ while assuming that all of the
objects have the same mass ratio $q$. The sum of $P_{obs}$ over all of
the objects is then the total expected number of observable sources
($N_{\rm obs}$). We discuss our results and their sensitivity to the
parameters in \S \ref{sec:results}.

%We mention here two important caveats.  First, 
A major uncertainty in the detectability comes from the size of the BLR,
$\rblr$.  As the the two BHs get closer than $\rblr$, the BLR will
envelope both of the BHs such that the broad emission lines are no
longer sensitive to the orbital motion of a single BH
\citep{2010shenloeb}. The BLR of the secondary BH could also suffer
tidal truncation at the Roche lobe surface as the binary gets tighter
\citep{2011montuori} and may be totally destroyed when the binary
separation is comparable to $\rblr$. Therefore, we take $\rblr$ as a lower 
limit of our integration in Eq. \ref{eq:rblr}. Our estimates of observability are very sensitive
to the assumed $\rblr$ because the smallest separations contribute most 
to the observability. A slightly larger $\rblr$ leads to a significant
decrease in our chances of detecting orbital motion of the
binaries. From reverberation mapping, the expected BLR size for a
single BH of $10^8 M_\odot$ (or $10^9 M_\odot$) is about 0.022 pc (or
0.07 pc) assuming the Eddington ratio of the QSO to be $\sim 0.1$
\citep{2010shenloeb}. 

We note, however, that if there is tidal truncation may change our
results.  Based on the locally optimally emitting cloud (LOC) model of
the BLR \citep{1995baldwin}, the {\mgii} emitting region is quite
broad \citep{2000korista}, and would be observable even in a BLR
one-tenth the fiducial assumed size.  We could be more sensitive to
SMBH binaries than the estimates in this section, if the BLR is more
compact in tight binaries. In the future, we will use the C{\tiny
  III}] line, which comes from a much more compact region \citep[][]{2005wayth}, to mitigate these uncertainties.

%%%%%%%%Observability at Fixed Separation %%%%%%%%

\subsection{Observability at Fixed Separation and Binary Mass Ratio}
\label{sec:fixaq}

Let us start with the observability of a BH binary $P_{obs}$ for a
given total mass $M_{\rm BH}$, mass ratio $q$, separation $r$, observing
time interval $\Delta t$ and orientation on the sky. Note that the
total mass $M_{\rm BH}$ is equal to $(1+q)/q$ times the BH mass shown in
Fig. \ref{fig:SampleInfo}d, since we assume that we are observing the
secondary BH. Although simulations show that very high eccentricity
may arise through binary-star interactions \citep[e.g.][]{2008sesana,
  2010sesana, 2011preto, 2012khan} and binary-disk interactions
\citep[e.g.][]{2005armitage, 2009cuadra, 2011roeding}, the
eccentricity of the binary orbit is highly uncertain. For simplicity, we assume zero
eccentricity throughout, so that the intrinsic velocity of the
secondary black hole is

\begin{equation}
V_0 = \frac{1}{1+q} \sqrt{\frac{GM_{\rm BH}}{r}}.
\end{equation}
The observed velocity is $V_{obs} = V_0 \sin\phi
\cos\theta$ where $\phi$ and $\theta$ are the azimuthal and polar
angle of the observer on the sphere centered on the black hole
binary. $\theta = 0$ when the binary orbit is edge-on relative to the
observer, and $\phi = 0$ when the secondary (accreting) BH lies
between the observer and the center of mass of the binary. If this
quasar is observed twice with a time interval of $\Delta t$, then the
drift in velocity of the secondary black hole is
\begin{eqnarray}
|\Delta V_{obs}| &=& V_0 \sin(\phi + \Omega \Delta t) \cos \theta - V_0 \sin\phi \cos\theta \nonumber \\
			&\approx& V_0 |\cos\theta| |\cos\phi| |\sin(\Omega\Delta t)|,
\end{eqnarray}
where $\Omega = \sqrt{GM_{\rm BH}/r^3}$ is the angular velocity.

If the observer has equal chances to be at any solid angle on this sphere,
then $\phi$ is uniformly distributed in $[0, 2\pi]$ and $|\cos\theta|$
also has an equal chance to have any value between $[0,1]$ according to
$d\Omega = \sin\theta d\theta d\phi = - d(\cos\theta) d\phi$. Let us
say that the smallest velocity drift we can measure is $\Delta V_{\rm
  lim}$ ($\Delta V_{\rm lim} = 340$ km~s$^{-1}$ for our whole sample and
$\Delta V_{\rm lim} = 280$ km~$^{-1}$ for the high 
subsample). We now calculate the probability that $|\Delta
V_{obs}|$ exceeds the observational limit.

For a given $\cos\theta$, the probability of $|\Delta V_{obs}| > \Delta V_{\rm lim}$, i.e. 
\begin{equation}
|\cos\phi| > \frac{\Delta V_{\rm lim} }{V_0 |\cos\theta| |\sin(\Omega \Delta t)|}
\end{equation}
in the range $\phi \in [0, 2\pi]$ is
\begin{equation}
\frac{2}{\pi} \arccos \left(\frac{\Delta V_{\rm lim} }{V_0 |\cos\theta|
  |\sin(\Omega \Delta t)|} \right).
\end{equation}
If we weight this value with the probability
distribution of $|\cos\theta|$ in the range $[0,1]$, the total
probability becomes
\begin{eqnarray}
&&P_{obs} (r, M_{\rm BH}, q, \Delta t) \nonumber \\
	 &=& \int_0^1 \frac{2}{\pi} \arccos \left( \frac{\Delta V_{\rm lim} }{V_0 |\cos\theta| |\sin(\Omega \Delta t)|} \right) d(|\cos \theta|)\nonumber \\
	&=& \frac{2}{\pi} \frac{\Delta V_{\rm lim}}{V_0 |\sin(\Omega \Delta t)|} \times \nonumber \\
	&&[ x \arccos(\frac{1}{x}) - \ln (x+\sqrt{x^2 -1}) ] \Big|_{x=1}^{x=V_0\sin(\Omega \Delta t) / \Delta V_{\rm lim}}.
\end{eqnarray}

\begin{figure*}[ht]
\centering
\includegraphics[width=0.8\textwidth]{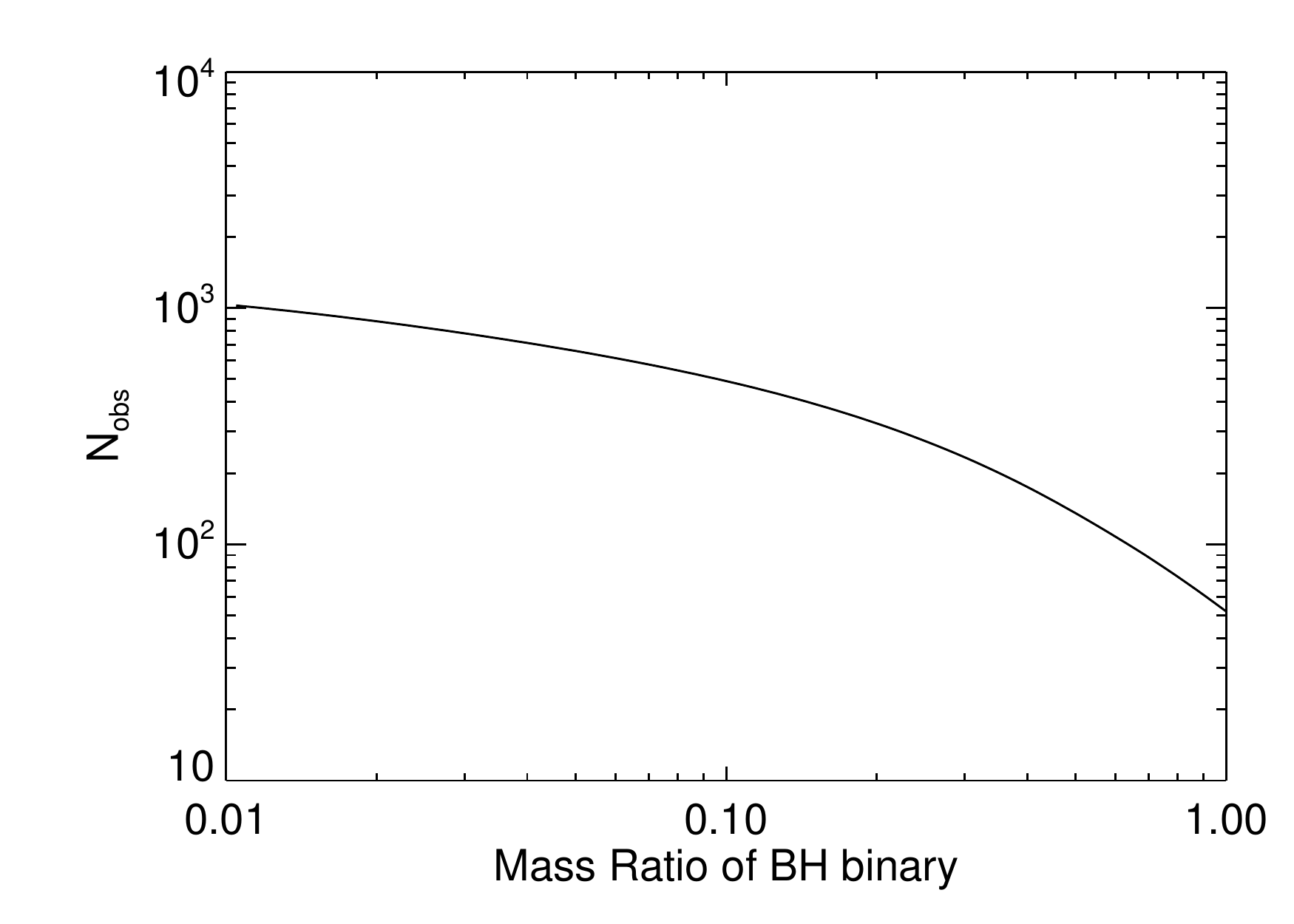}
\caption{Dependence of expected number of observable binaries ($N_{\rm obs}$) on 
the mass ratio $q$ of SMBH binaries at a fixed separation of 0.1 pc. The sensitivity mainly 
comes from the dependence of total BH mass on $q$, given an observed secondary BH mass.}
\label{fig:expectratio}
\end{figure*}

\begin{figure*}[ht]
	\centering
	\includegraphics[width=0.47\textwidth]{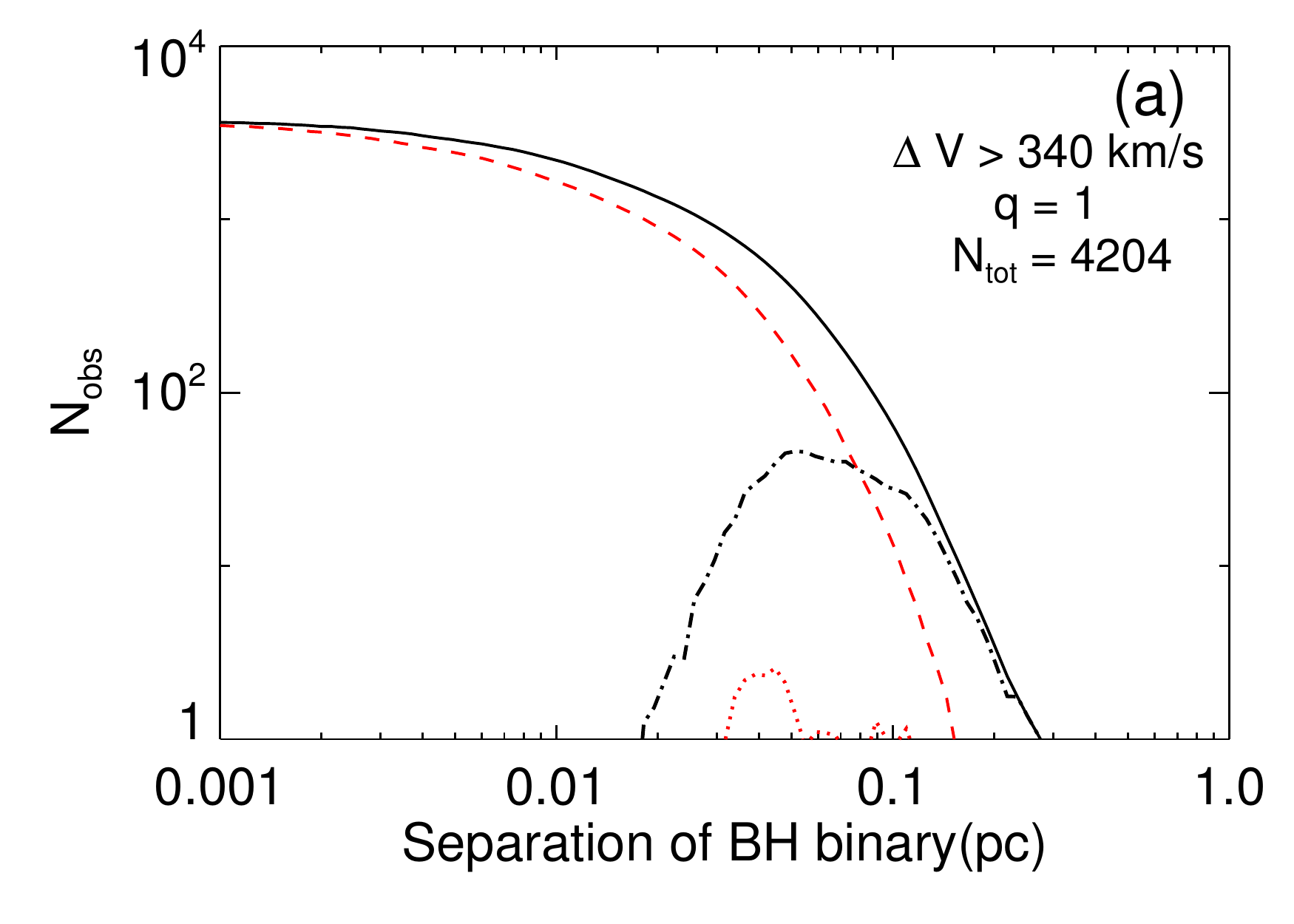}
	\includegraphics[width=0.47\textwidth]{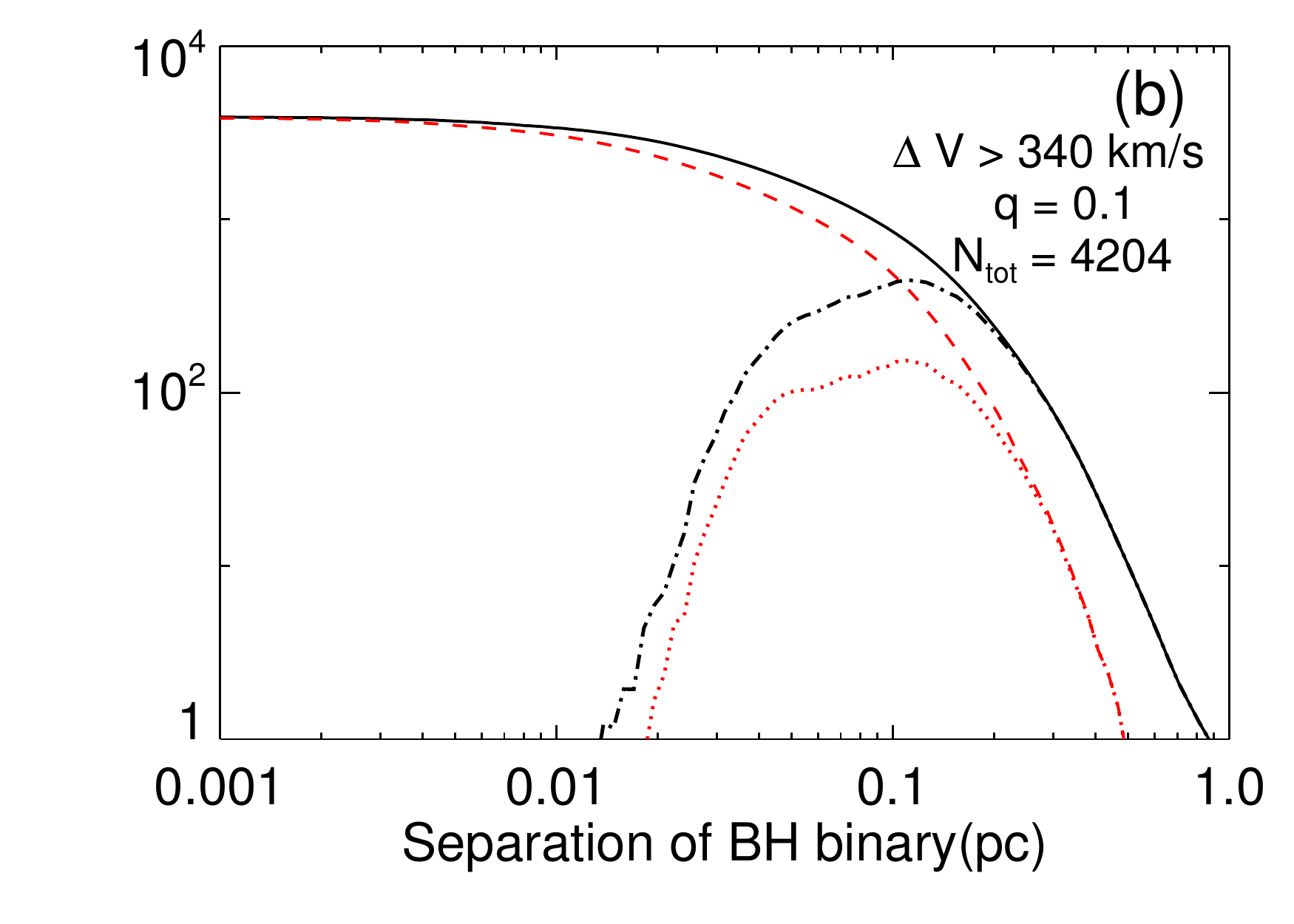}
	\includegraphics[width=0.47\textwidth]{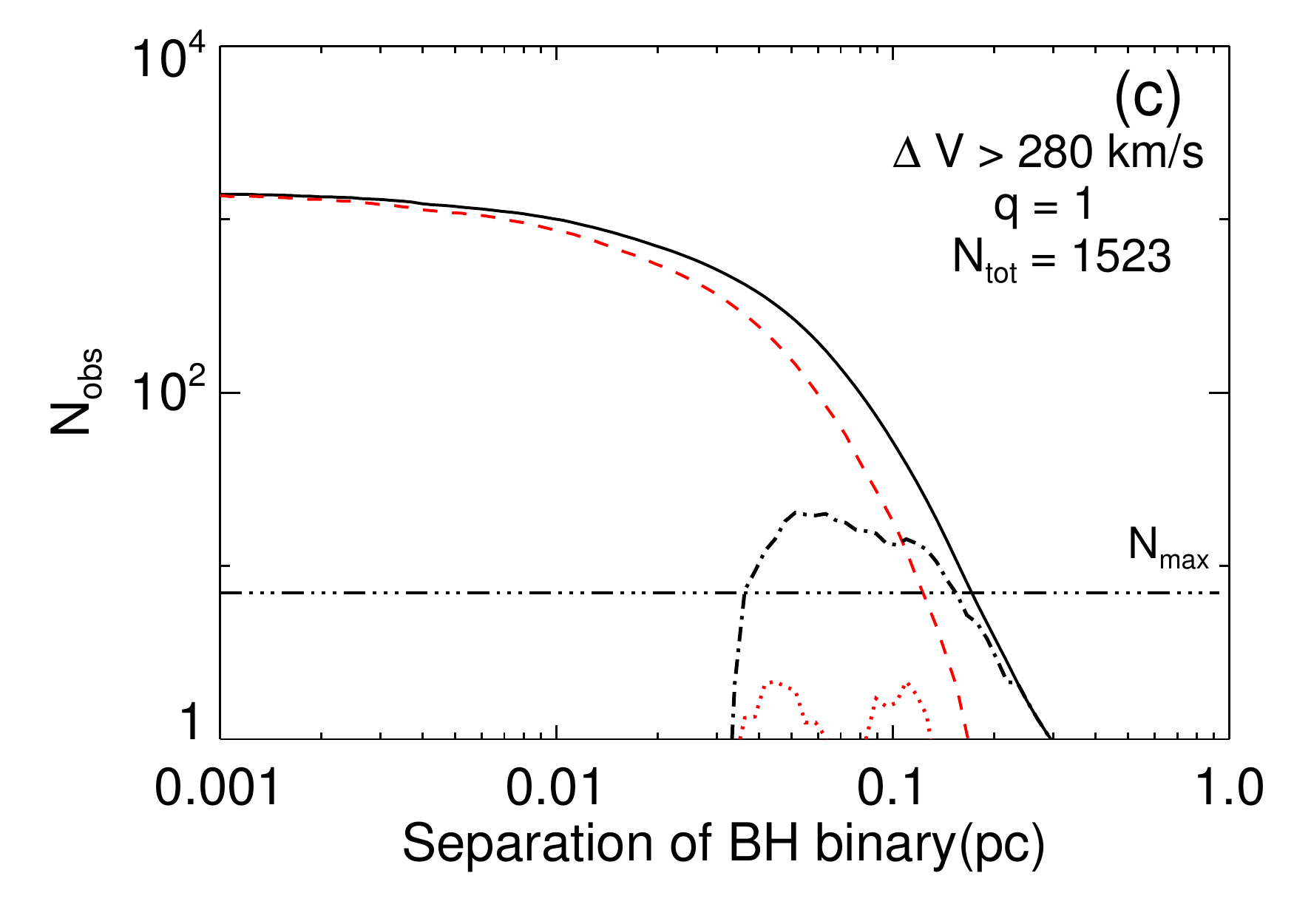}
	\includegraphics[width=0.47\textwidth]{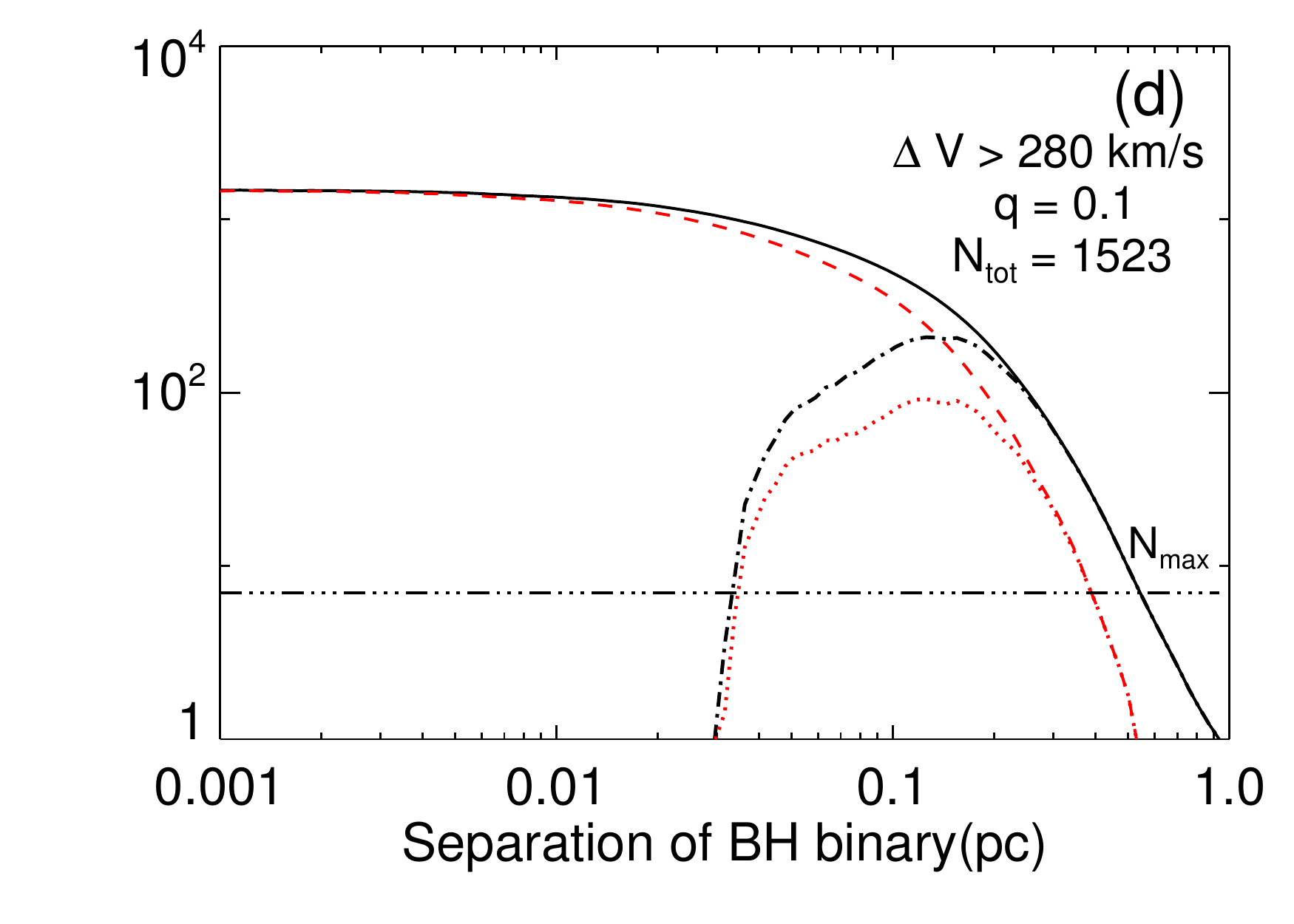}
\caption{
  Dependence of the expected number of candidates ($N_{\rm obs}$) on the
  separation of the SMBH binaries. (a) and (b) are for the whole
  sample with $N_{tot} = 4204$, and (c) and (d) are for the S/N/pixel $> 10$
  subsample with $N_{tot} = 1523$.  The black (solid and dash-dotted)
  lines utilizes BH masses from broad-line width measurements, while
  the red (dashed and dotted) lines use the lower-limits to the BH
  masses as estimated from the bolometric luminosity assuming an
  Eddington ratio $\dot{m}_{Edd}=0.5$. The dash-dotted and dotted
  lines include truncation at $\rblr$ in the calculations, while the
  solid and dashed lines do not. The horizontal lines in the lower
  panel are the upper limits from this experiment ($N \leq 7$).}
\label{fig:expect}
\end{figure*}

We calculate the observability above for each object in our sample and
sum them, which gives us the total number of candidates.
The dependence of the total expected number of candidates on the mass
ratio and separation of the BH binary are shown in
Fig. \ref{fig:expectratio} and \ref{fig:expect} respectively.  In
Fig. \ref{fig:expectratio} we fix the separation to be $0.1$ pc and
vary the mass ratio $q$ from 0.01 to 1. The observability is
moderately sensitive to the mass ratio. The expected number of
observable sources increases by a factor of 10 as $q$ decreases by a
factor of 10. This sensitivity is mainly due to the dependence of the
total BH mass $M_{\rm BH}$ on $q$, as we assume that we observe the
secondary BH. The number of observable candidates increases more
steeply as the radius decreases (Fig. \ref{fig:expect}), because the
orbital velocity of the BHs is proportional to $r^{-1/2}$.  No BH
binaries are observable if the separation gets larger than $\sim 0.2$
pc for $q=1$ (or $\sim 0.5$ pc if $q=0.1$). We do additional
calculations at fixed separation while truncating the observability at
$\rblr$ for each object (dash-dotted and dotted lines in
Fig. \ref{fig:expect}). In \S \ref{sec:results} we will compare our
number of candidates with the models in the scenario where all
quasars live at a fixed separation forever.

%% f5

%% f6

%%%%%%%% Time Evolution of the SMBH Binary %%%%%%%%%%%%%
\subsection{Time Evolution of the SMBH Binary}
\label{sec:timeevolution}

We now calculate the second ingredient needed to compute $P_{obs}$ using
equation (\ref{eq:rblr}) -- the residence time $t_{\rm res}$ as a function 
of the binary separation $r$ for a given set of binary characteristics.

From Table \ref{table:candidates}, our
candidate SMBH binaries are at separations of $0.01 \sim 0.1$ pc or
closer, well inside the regime of ``the final parsec''. As mentioned in the Introduction, there are many mechanisms proposed
to overcome the final parsec problem. One of the most promising
is tidal interaction of the binary with a circumbinary
disk, naturally leading to the orbital decay of the binary, which
loses its angular momentum to the disk \citep[e.g.][]{2009haiman,
  2009cuadra, 2012rafikov}. It is generally expected that a massive
binary clears a cavity in a disk around itself so that the mass inflow
onto its components is considerably suppressed compared to the mass
accretion rate $\dot M_\infty$ far out in the disk. Mass accumulation
at the inner edge of the disk caused by the tidal barrier modifies the
disk structure over time and accelerates binary inspiral.

In this work we compute $t_{\rm res}$ in the framework of this 
{\it gas-assisted} binary inspiral scenario. If we assume that all
quasars are triggered by gas accretion at $\sim 0.1$pc and the
inspiral of the central SMBH binaries is dominated by the
interaction with a gas disk, then the probability of observing a pair
at a given radius is just proportional to the {\it residence time}
$t_{\rm res}\equiv |d\ln r/dt)|^{-1}$.  This time is obviously a function
of the physical mechanism driving the orbital evolution of the
binary. As a result, our observations in principle provide a way of
constraining different inspiral mechanisms.

%% f7
\begin{figure*}[!htbp]
	\centering
	\includegraphics[width=0.49\textwidth]{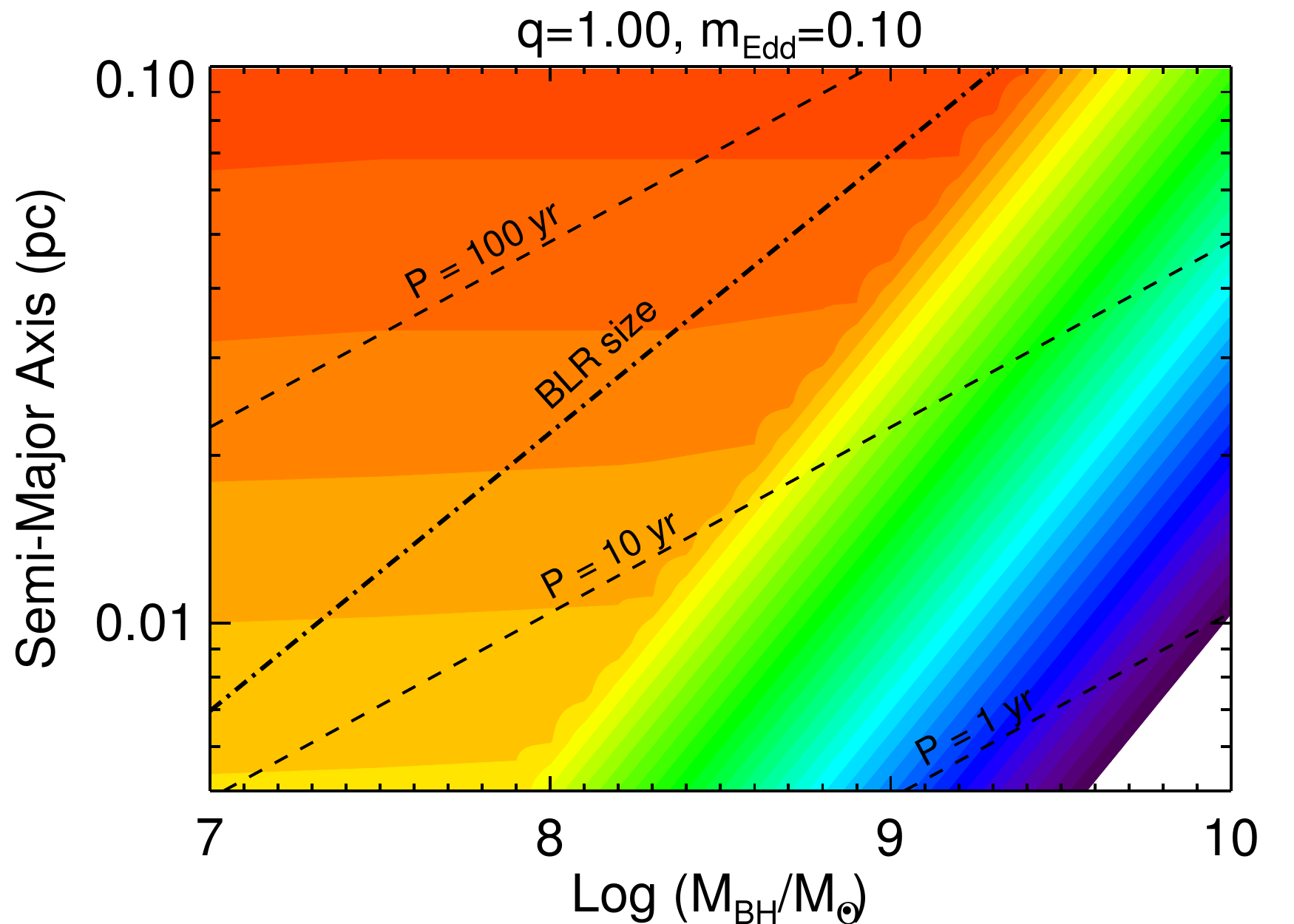}
	\includegraphics[width=0.49\textwidth]{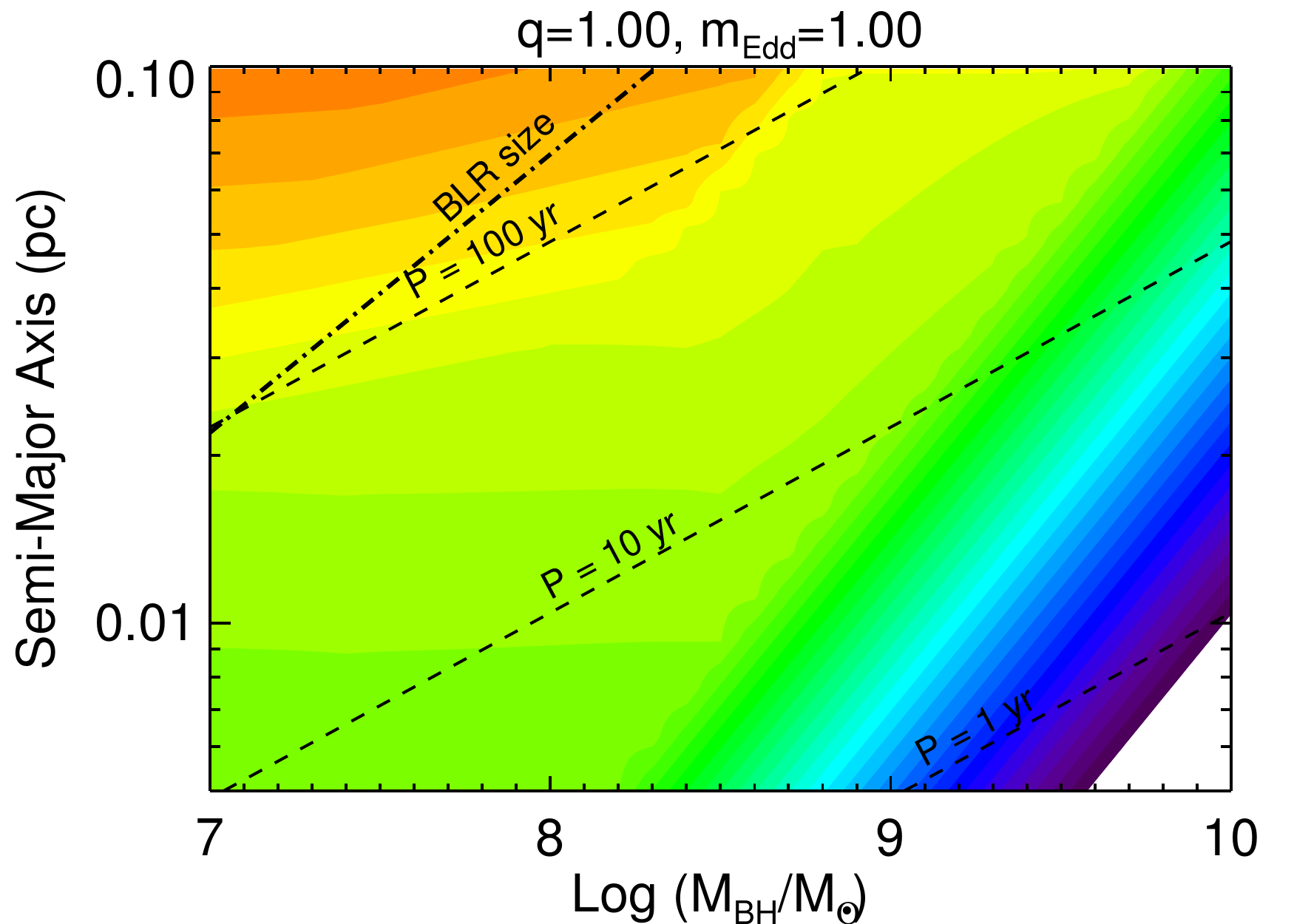}
	\includegraphics[width=0.49\textwidth]{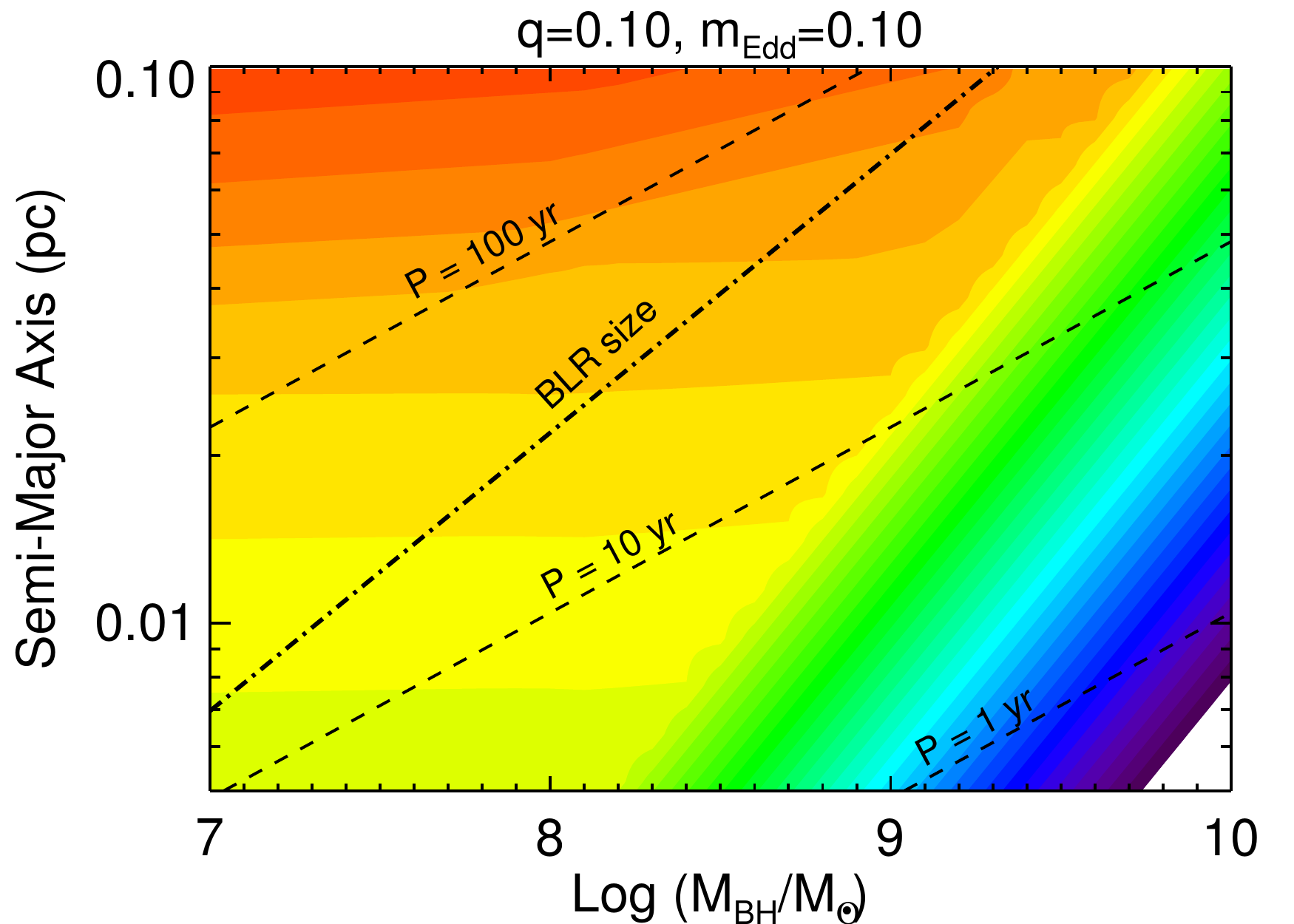}
	\includegraphics[width=0.49\textwidth]{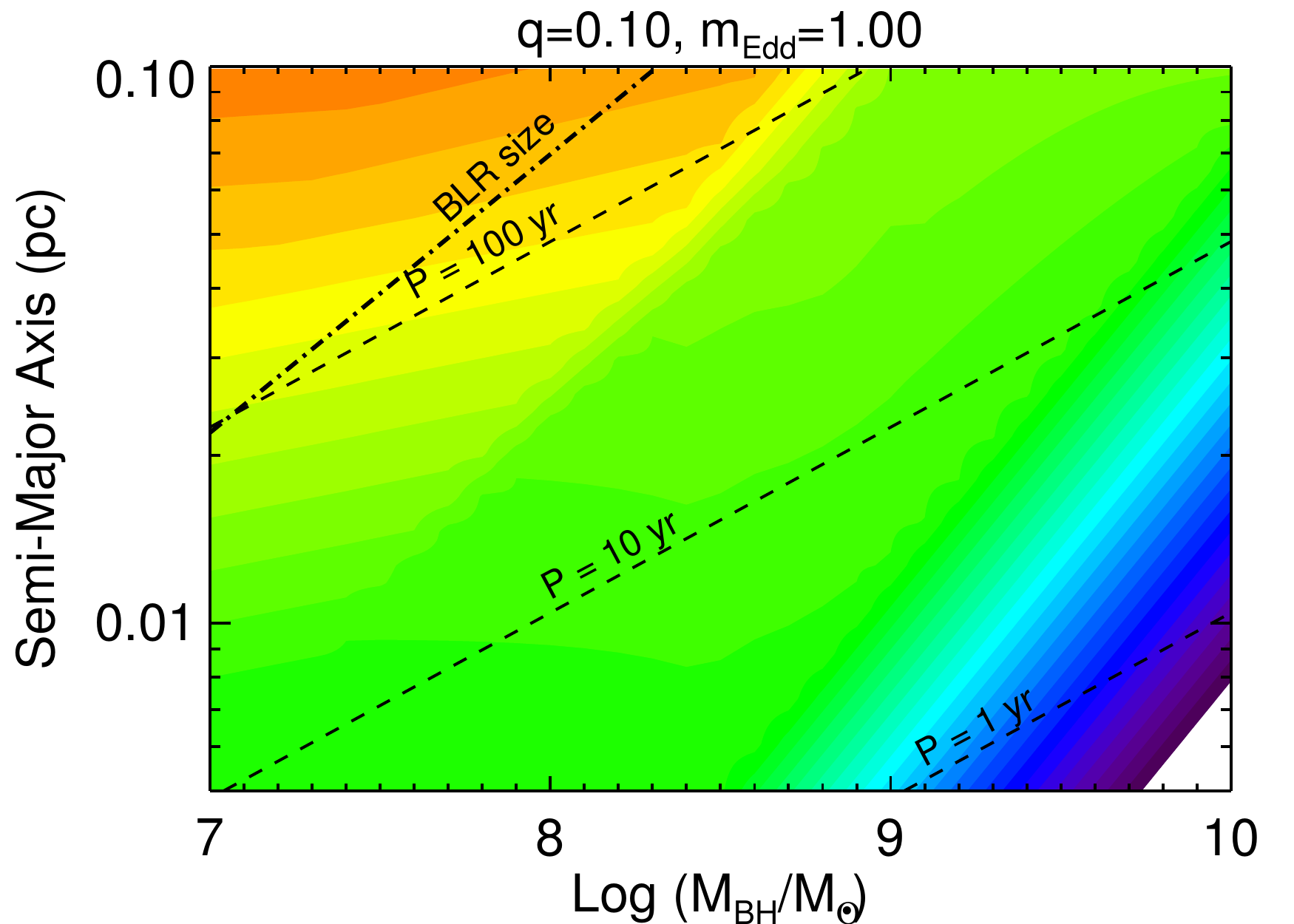}
	\includegraphics[width=0.49\textwidth]{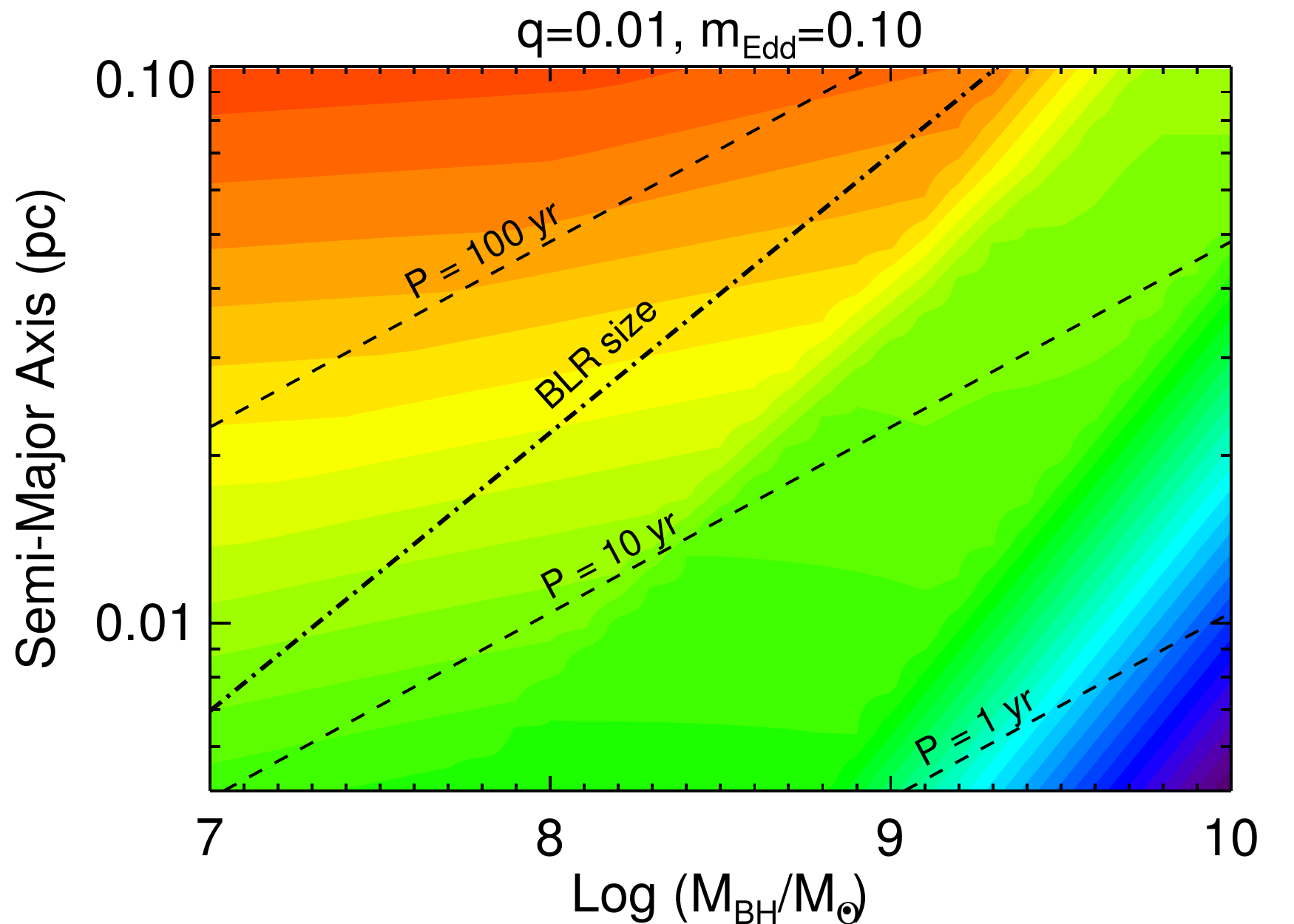}
	\includegraphics[width=0.49\textwidth]{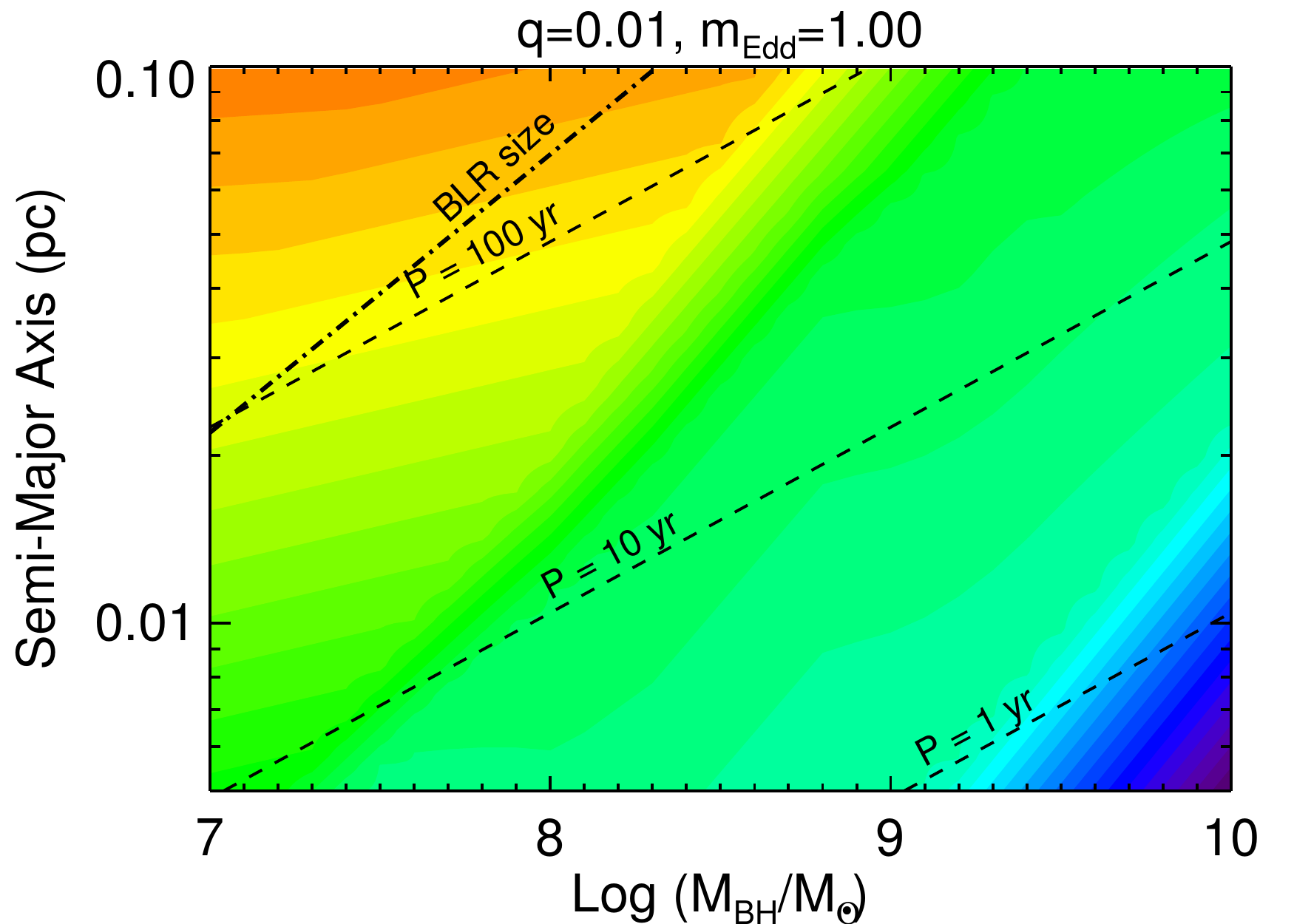}
\begin{center}
	\centering
	\includegraphics[width=0.5\textwidth]{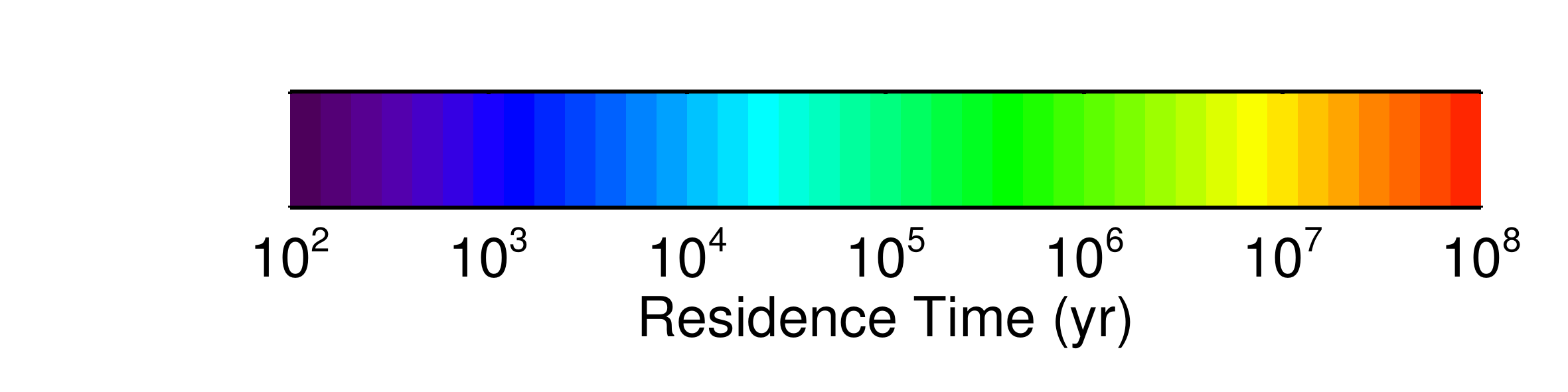}
\end{center}
\caption{
  SMBH binary evolution driven by gas disks: color maps of residence
  time $t_{\rm res}$ at various radii $r$ for a range of BH masses. The
value of $t_{\rm res}$ is indicated by the color bar. Plots on the left 
(right) assume an Eddington accretion rate ratio ($\dot m_{\rm Edd}$) of 0.1
  (1.0). Plots on the top, center and bottom assume that the mass ratio of the
  BH binary is 1, 0.1, 0.01 respectively. Dashed lines are curves of 
constant orbital period (labeled), dot-dashed line is the $\rblr$
given by equation (\ref{eq:BLR}). See text for more details.}
\label{fig:tres}
\end{figure*}

Our calculation of $t_{\rm res}$ closely follows that in
\citet{2012rafikov}, which provides a recipe for computing the fully coupled, 
time-dependent binary+disk evolution, and should be referred to
for details. We start binaries of all masses at a semi-major axis of
$0.1$ pc, in a circumbinary disk with viscosity characterized by
$\alpha=0.1$. We assume that in a radiation pressure-dominated regime
(which is valid at small separations) viscosity in the disk is
proportional to the total (rather than the gas) pressure. The value of
the mass accretion rate in the disk far from the binary $\dot
M_\infty$ is normalized by the Eddington rate $\dot M_{\rm Edd}$ computed
for the full binary mass and radiative efficiency of
$\varepsilon=0.1$. Evolutionary calculations are always started with a
constant $\dot M=\dot M_\infty$ disk extending all the way into the
initial binary orbit at $0.1$ pc.  Such an initial setup naturally 
arises if the gas that has fallen into the center of the galaxy 
circularizes outside the binary orbit and then viscously spreads 
inwards. For simplicity our calculations ignore the issue of 
gravitational stability of the circumbinary disk at large separations,
which is poorly understood at the moment \citep{2003goodman, 2009rafikov}.

\citet{2012rafikov} has shown that the torque acting on the binary 
and driving its inspiral is set by $\dot M_\infty$, the 
radius of influence $r_{\rm infl}$ out to which the binary 
torque has been propagated by the viscous stresses in the disk, and 
the degree of mass overflow across the orbit of the 
secondary, i.e. the fraction $\chi< 1$ of $\dot M_\infty$ that 
penetrates into the cavity and gets accreted by the binary 
components. The dependence on the latter is rather weak and
the orbital evolution is similar to the no-overflow case as long as 
$\chi\lesssim 0.2$, which we assume to be the case in this work 
for simplicity. Of course, in practice some overflow is needed
to power the observed quasar activity, so $\chi$ is not strictly zero.

The time dependence of the radius of influence is self-consistently
computed by following the viscous evolution of the circumbinary 
disk torqued by the binary, as described in \citet{2012rafikov}. 
In calculating $r_{\rm infl}$ we 
keep track of the different physical regimes in the disk at
$r_{\rm infl}$. This is important since $r_{\rm infl}$ can 
vary by orders of magnitude with respect to the binary 
semi-major axis in the course of its inspiral, 
allowing  the opacity behavior and relative role of radiation 
pressure at $r_{\rm infl}$ to change over time.
This self-consistent calculation favorably distinguishes our present calculation (and 
that in \citealt{2012rafikov}) from the self-similar solution of 
\citet{1999ivanov} or the SMBH inspiral calculations in 
\citet{2009haiman}. 

One particular aspect of the $t_{\rm res}$ calculation that we wish 
to emphasize is that in many cases we find the system in so-called 
{\it disk-dominated} evolution, a regime when the ``local disk mass''\footnote{Definitions of $M_d$
different by factors of $2\pi$ or $4\pi$ can be found in the literature.}  
$M_d=\Sigma r^2$ at the inner edge of the disk (here $\Sigma$ is the 
local surface density of the disk enhanced by the mass pileup near 
the cavity edge) is larger than
the mass of the secondary $M_s$. This regime is often valid at 
the initial stages of evolution, especially for lower SMBH 
masses, because of the large starting semi-major axis of 0.1 pc 
that we adopt in this work. In this regime we follow conventional
wisdom \citep{1997ward, 2009haiman} and simply assume that 
the secondary passively follows viscous evolution of the disk,
which is equivalent to setting $t_{\rm res}=t_\nu$, where 
$t_\nu=r^2/\nu$ is the viscous time at the inner edge of the 
circumbinary disk, computed using the {\it local} value of viscosity 
$\nu$ and accounting for the possibility of mass pileup at the inner 
disk edge. As the binary orbit shrinks, the system
inevitably transitions into the {\it secondary-dominated} regime, 
when $M_s>M_d$. As soon as this happens we let the disk respond
viscously to the binary torque. On the other hand, \citet{2012rafikov}
has shown that deep in the secondary-dominated regime ($M_d\ll M_s$) 
the evolution (or residence) time is given by 
\ba
t_{\rm res}=t_\nu\frac{1}{6\pi(1+q)}\frac{M_s}{M_d}.
\label{eq:r_relation}
\ea
Thus, $t_{\rm res}$ evaluated according to this formula becomes
equal to $t_\nu$ only when $M_d= M_s/[6\pi(1+q)]$,
which is significantly less than $M_s$. The precise details of
the transition between the disk- and secondary-dominated regimes
are not well understood at the moment, so in this 
work we assume for simplicity that $t_{\rm res}=t_\nu$ for 
$M_d>M_s/[6\pi(1+q)]$, and switches to the behavior predicted 
by equation (\ref{eq:r_relation}) only for $M_d< M_s/[6\pi(1+q)]$. 
This assumption makes the viscous evolution of the disk quite important
and can potentially result in an overestimate of $t_{\rm res}$ at 
intermediate separations ($\sim 0.01-0.1$ pc) by a factor of 
several. We cannot be more accurate in this regime without
better understanding the regime when $M_d\sim M_s$.

In Figure \ref{fig:tres} we show maps of $t_{\rm res}$ computed 
according to the aforementioned prescription \citep{2012rafikov} for 
several values of $q$ and $\dot M_\infty/\dot M_{\rm Edd}$. 
Several features of these maps are worth discussing. At high 
masses (and small separations) color contours turn into straight 
lines which reflects the fact that in this regime the binary 
inspiral is determined by gravitational wave (GW) emission.
In this case $t_{\rm res}$ is simply given by  
\ba
t_{\rm GW}(r_b)=\frac{5(1+q)^2}{8q}\frac{R_S}{c}
\left(\frac{r_b}{R_S}\right)^4,
\label{eq:t_GW}
\ea
where $R_S\equiv 2GM_{\rm BH}/c^2$ is the Schwarzschild radius 
of the black hole with the combined mass $M_{\rm BH}$. This timescale
is very sensitive to $M_{\rm BH}$ and $r$, as can be seen in 
Figure \ref{fig:tres}. However, the quasars used in our study
almost never probe this part of the parameter space because the 
size of the broad line region $\rblr$ (shown 
by the dot-dashed line in $t_{\rm res}$ maps) exceeds the maximum 
separation at which GW emission dominates for all masses. Thus, 
our quasars are in the regime when the disk torque 
dominates binary inspiral.

Figure \ref{fig:tres} also reveals some obvious trends with $q$
and $\dot m_{\rm Edd}$. Lower $q$ results in faster 
inspiral since for a lower-mass secondary the binary contains 
less angular momentum but the disk torque is relatively insensitive 
to $q$, see \citet{2012rafikov}. Higher $\dot m_{\rm Edd}$ at a
fixed $M_{\rm BH}$ implies a circumbinary disk with higher surface density, 
which provides more torque, considerably accelerating the orbital 
evolution of the binary.  Higher $\dot m_{\rm Edd}$ also extends the disk-dominated 
phase of the binary evolution. In particular, at low masses 
(and large separations) $t_{\rm res}$ exhibits rather 
weak dependence on $M_{\rm BH}$, which is a result of the system
evolving in a disk-dominated regime where the inspiral time is
given by the local viscous timescale, which is only weakly 
dependent on $M_{\rm BH}$. 

Quite important for our work is the relatively weak dependence of
$t_{\rm res}$ on separation in some cases, e.g. low $\dot m_{\rm Edd}$ and
relatively high $q$, which results in rather long $t_{\rm res}$ for $r\sim
0.003-0.1$ pc and increases the number of binaries in this range of
$r$. This behavior is a result of two main factors. First, initially,
when the system is in the disk-dominated regime, the inner parts of
the disk around the binary are gas-pressure dominated. Under these
conditions the viscous timescale on which the binary orbit evolves in
the disk-dominated case is a relatively weak function of $r$ compared
to the radiation pressure-dominated case typical for high $\dot m_{\rm
  Edd}$ \citep{2012rafikov}, so that $t_{\rm res}$ is quite long at
$r\gtrsim 0.01$ pc.

Second, even after evolution switches to the
secondary-dominated regime, binary inspiral is slow
and is still not very sensitive to the separation [e.g. see
the behavior of $t_{\rm res}$ in Fig. 8 of \citet{2012rafikov}].
It is very important that here we self-consistently follow the
time-dependent structure of the whole disk in the spirit of
\citet{1999ivanov}. Previously, \citet{2009haiman} found
considerably faster binary evolution (lower $t_{\rm res}$) for
the same $M_{\rm BH}$, $r$, etc.\ when employing the quasi-steady
state circumbinary disk solutions derived in \citet{1995syer},
which do not apply to real disks, see
\citet{1999ivanov,2012rafikov} for details. The
non-locality of the binary-disk coupling in the
secondary-dominated regime \citep{2012rafikov} ---
the fact that for a given value of $\dot M$ at the inner disk edge 
(equal to zero in the case of no overflow) the magnitude of the torque 
acting on the binary is set by the
disk properties not at its inner edge, which is often
radiation-pressure dominated, but much further out,
at $r_{\rm infl}$, where the disk is gas-pressure dominated
--- is also very important for determining the residence time.

%%%%%%%%%%%%%% Results %%%%%%%%%%%%%%%%%%%%%%%%%

\subsection{Results: Expected Number of Observable SMBH Binaries}
\label{sec:results}

In computing observability, we consider two basic scenarios for the BH
binary: no temporal evolution, and radial migration via a gaseous
accretion disk.  From each, combined with our upper limit of seven
candidates, we can infer something about the BH binary population in
luminous $z \approx 1-2$ QSOs.  In the first scenario, all SMBH
binaries live at fixed separation with no time evolution as we
discussed in \S \ref{sec:fixaq}. The expected numbers of observable
sources from our high S/N subsample versus binary separation is
shown in Fig. \ref{fig:expect}. Comparison of this theoretical
distribution with our observational upper limit ($N \le 8$) rules out
that all the $\sim 10^9 M_\odot$ BHs that we observe at $z \approx
1-2$ exist in binaries in the separation range of $\sim [0.03, 0.2]$
pc. Binaries closer than 0.03 pc are not observable due to the
truncation or destruction of the BLR, while binaries with $r > 0.2$ pc are
not observable given the $\Delta t$ distribution of our sample.  We
note that even this basic result is of cosmological interest.  For
instance, in the merger-tree models of Volonteri et al. (2003), many
major mergers are expected in the massive halos that host our massive
BHs between $1 < z < 2$.  We can rule out that the lifetime of the
resulting tight binaries is comparable to the Hubble time in all cases
(\S \ref{sec:summary}). 

In the second scenario, the binaries evolve through a gas disk.  With
the residence time $t_{\rm res}(r, M_{\rm BH}, \dot{m}_{\rm Edd})$ calculated
in the previous section (\S \ref{sec:timeevolution},
Fig. \ref{fig:tres}), the probability $P_{obs}(M_{\rm BH}, q, \Delta t,
\dot{m}_{\rm Edd})$ is evaluated according to Eq. \ref{eq:rblr} for
each object in our sample using its observed $M_{\rm BH}$ and $\Delta t$
and assuming that all of the objects have the same mass ratio $q$. The
sum of $P_{obs}$ over the objects is then the total expected number of
observable sources ($N_{\rm obs}$). As we discussed in \S
\ref{sec:sample}, the virial masses of the secondary BHs shown in
Fig. \ref{fig:SampleInfo}d imply an $\dot{m}_{\rm Edd}$ distribution
peaked at $\sim$ 0.25. For the first calculation of $N_{\rm obs}$, we use
$t_{\rm res}$ in the case that $\dot{m}_{\rm Edd} \sim 0.1$ and the virial
BH masses. Then, to bracket the known large uncertainties in $M_{\rm BH}$
measurements \citep[e.g.,][]{2006vestergaardpeterson, 2008shen},
we also do the calculations assuming that $\dot{m}_{\rm Edd}=1$ and
adopting $M_{\rm BH}$ accordingly. In both of the calculations, we
truncate the binary migration at $\rblr$ for each object according to
Eq. \ref{eq:rblr}, using the virial and Eddington-limited BH masses
respectively. We show the range of expected $N_{\rm obs}$ in our S/N/pixel $>10$
subsample for $\dot{m}_{\rm Edd}=0.1, 1$ and $q=1, 0.1$ in Table
\ref{table:results}.

If we adopt the most likely parameters $\dot{m}_{\rm Edd}=0.1$ and
$q=1$, the final expected number of SMBH binaries in the high S/N
subsample is $N_{obs} = 24$. Comparison of this expectation with our
seven candidates implies that $<30\%$ of quasars host SMBH binaries
with $r < 0.1$ pc. A more extreme mass ratio of $q=0.1$, which seems
unlikely for the bulk of our sample, increases the expected $N_{\rm obs}$
by a factor of $\sim 8.5$.

Note, however, that our results are very sensitive to the extrapolated
$\rblr - L_{\rm bol}$ scaling relation, which is highly
uncertain. Since our method is only sensitive to binaries that are
less than $\sim$ 0.1 pc apart (see Fig. \ref{fig:expect}), the final
expected number of observable candidates is actually determined by the
detectability within the range $[ \rblr$, 0.1 pc$]$. Candidates in
this range are the most massive BHs with the corresponding largest
velocity drifts, so their BLR size is approaching $\sim 0.1$ pc (e.g.,
$\rblr \sim$ 0.07pc for $10^9 M_\odot$ assuming $\dot{m}_{\rm
  Edd}=0.1$). A slightly larger $\rblr$ makes the valid range $[
\rblr$, 0.1 pc$]$ much narrower. We will test the sensitivity to
$\rblr$ quantitatively in \S \ref{sec:next}.
%For example, if we increase the assumed Eddington ratio to $\dot{m}_{\rm Edd}=1$, which increases $\rblr$ by a factor of $\sqrt{10}$ (Eq. \ref{eq:BLR}), then we expect $zero$ detection. 

The sensitivity of our results to $\dot m_{\rm Edd}$ has two
origins. First, the BH masses we use scale with $\dot m_{\rm
  Edd}$ given the observed luminosity. Second, the
gas-assisted evolution of BH binaries depends on $\dot m_{\rm Edd}$:
in the lower $\dot m_{\rm Edd}$ case the circumbinary disk is less
massive meaning (1) more significant dominance of the secondary BH
over the local disk (larger $M_s/M_d$) and (2) less torque acting on
the binary (which is proportional to $\dot M_\infty\propto \dot m_{\rm
  Edd}$). These factors increase $t_{\rm res}$ for $\rblr<r< 0.1$ pc in
our scenario of the gas-assisted evolution, boosting the expected
$N_{\rm obs}$ for lower $\dot m_{\rm Edd}$.

Our results are also very sensitive to the details of the gas-assisted
evolution model in the intermediate separation range $[ \rblr$, 0.1
pc$]$. Evolutionary models where BH binaries spend a longer time in this
radial range predict a larger number of observable sources $N_{\rm obs}$. In our
calculation of $t_{\rm res}$ following \citet{2012rafikov} in \S
\ref{sec:timeevolution}, this separation range falls in the
secondary-dominated regime, in which matter piles up outside the
binary orbit, binary evolution is slow and is described by
Eq. \ref{eq:r_relation}. Faster orbital evolution in this separation
range, as suggested by \citet{2009haiman} based on the circumbinary disk
solution by \citet{1995syer}, would result in a smaller number of
detectable SMBH binaries. Thus, it should in principle be possible in
the future to discriminate between the different evolutionary
scenarios.

Another important mechanism that may solve the final parsec problem is
that stars in a triaxial galaxy nucleus could continuously replenish
the loss cone and drive subsequent inspiral evolution
\citep{2004merrittpoon, 2011merrittvas}. \citet{2004merrittpoon} found
that stars on chaotic (centrophilic) orbits in a triaxial galactic
nucleus could refill the loss cone and feed the central BHs at quite a
high rate and thus drive the binary evolution at sub-parsec
scales. \citet{2011merrittvas} found that there exists a family of
stellar orbits in triaxial galaxies called pyramids orbits. The
angular momentum of stars can reach zero at the ``corners'' of the
pyramid, at which point they are easily captured by the central
BHs. The lifetime of the binary then becomes a function of the
structure of the merged remnant \citep{2012khan} but may take billions
of years. In this scenario, quasars could be ignited at any radius,
since the lifetime of the quasar is likely much shorter than the
inspiral timescale of the BH binaries. However, without more
assumptions about the dynamical state of the galaxy centers to
determine $t_{\rm res}$, we cannot easily predict the detectability of
SMBH binaries in this scenario.

In the most pessimistic evolutionary scenario, the SMBH binaries stop
spiraling inwards due to depletion of stars and stall there
forever. \citet{1996quinlan} did scattering experiments and found that
the orbital decay of the binary stalls when most of the stars have
been ejected, and this stalling separation is 0.01pc $-$ 1pc depending
on the mass and mass ratio of the BH binary. The stalling separation
we extract from \citet[see their Fig. 7]{1996quinlan} is

\begin{eqnarray}
&&\log_{10}\left(\frac{R_{stall}}{pc}\right) \nonumber \\
&=& 0.082 \log_{10}\left(\frac{M_{\rm BH}}{M_\odot}\right) + 0.448 \log_{10} q - 0.474.
\label{eq:rstall}
\end{eqnarray}

For major mergers ($q \sim 1$), the stalling radius is around $\sim$ 1
pc. If there is no subsequent driving mechanism and all binary BHs
stall at $\sim$ 1 pc, our experiment would not be sensitive to any
binary BHs according to Fig. \ref{fig:expect}. With time baselines of only 
a few years, as probed here, we cannot hope to see radial velocity changes in 
pc-scale binaries. However, as we will see in the next section, next generation
surveys could significantly increase the observability even in this
scenario due to longer observing time intervals.

%%%%%%%%%%%%%%%%%%%%%%%%%%%%%%%%%%%%%%%%%%%%%%%%%%%%%%%%%%%%%%%%%%
%%%%%%%%%%%%%%%% Next Generation Surveys %%%%%%%%%%%%%%%%%%%%%%%%%%%%%%%
%%%%%%%%%%%%%%%%%%%%%%%%%%%%%%%%%%%%%%%%%%%%%%%%%%%%%%%%%%%%%%%%%%

\section{Next Generation Surveys}
\label{sec:next}

As part of the Baryon Oscillation Spectroscopic Survey (BOSS,
\citealt{2009schlegel}; \citealt{2013dawson}), 14,235 SDSS quasars,
mostly in the redshift range $2.1< z < 3.0$, are being revisited with
$\sim 10$-year time separations, with increasing numbers as the survey
goes on. This provides a great opportunity to search for SMBH binaries
using our cross-correlation methodology. We can use either C {\tiny III}] 
or C~{\tiny IV} to measure shifts in this sample. As discussed above,
C~{\tiny IV} may show many variations in line-shape that have nothing
to do with a binary companion, but arise (for instance) in a disk wind
\citep[e.g.][]{2000proga}. We believe C {\tiny III}]  should be quite promising 
however. To investigate the potential impact of the
BOSS multi-epoch sample, as well as future surveys, we predict the observability
of BH binaries as a function of time interval. Since the BH masses
measured from C~{\tiny IV} widths (currently available for 9,002 of the
14,235 quasars, \citealt{2011shen2}) are highly uncertain, we simply
adopt the same mass distribution as our DR7 sample for the purpose of
illustration.

\begin{figure*}[!ht]
	\centering
	\includegraphics[width=0.42\textwidth]{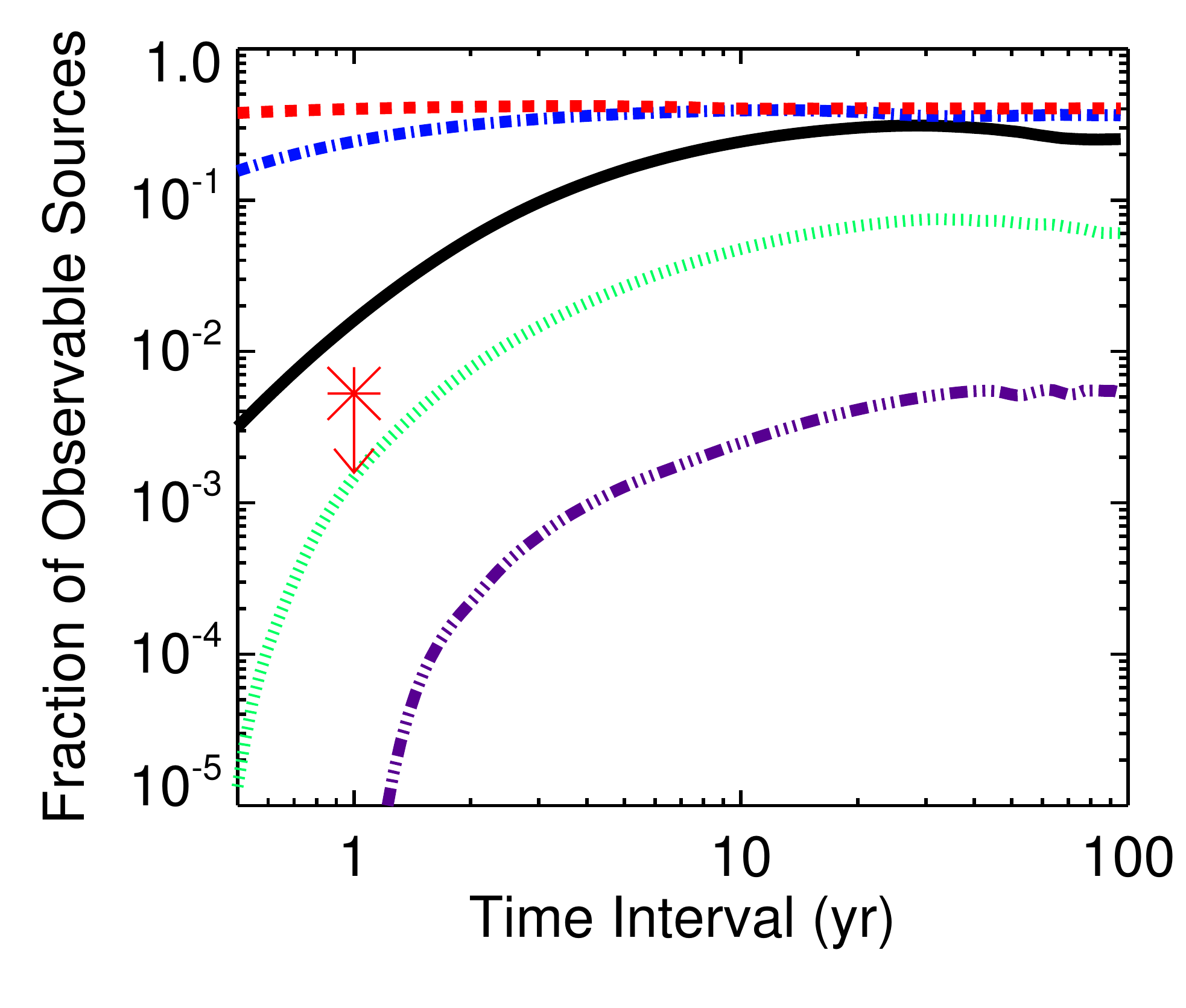}
	\includegraphics[width=0.42\textwidth]{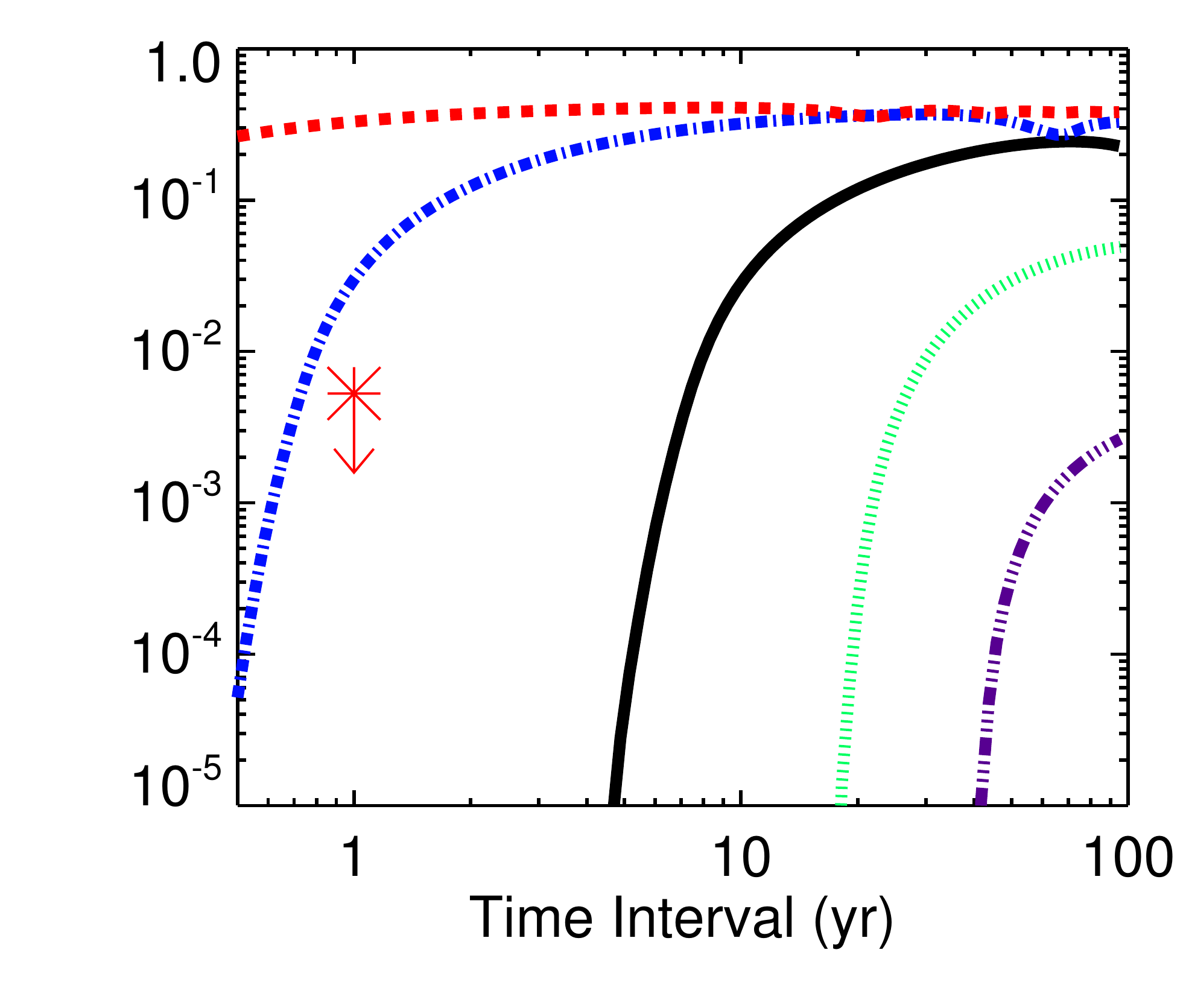}
	\includegraphics[width=0.13\textwidth]{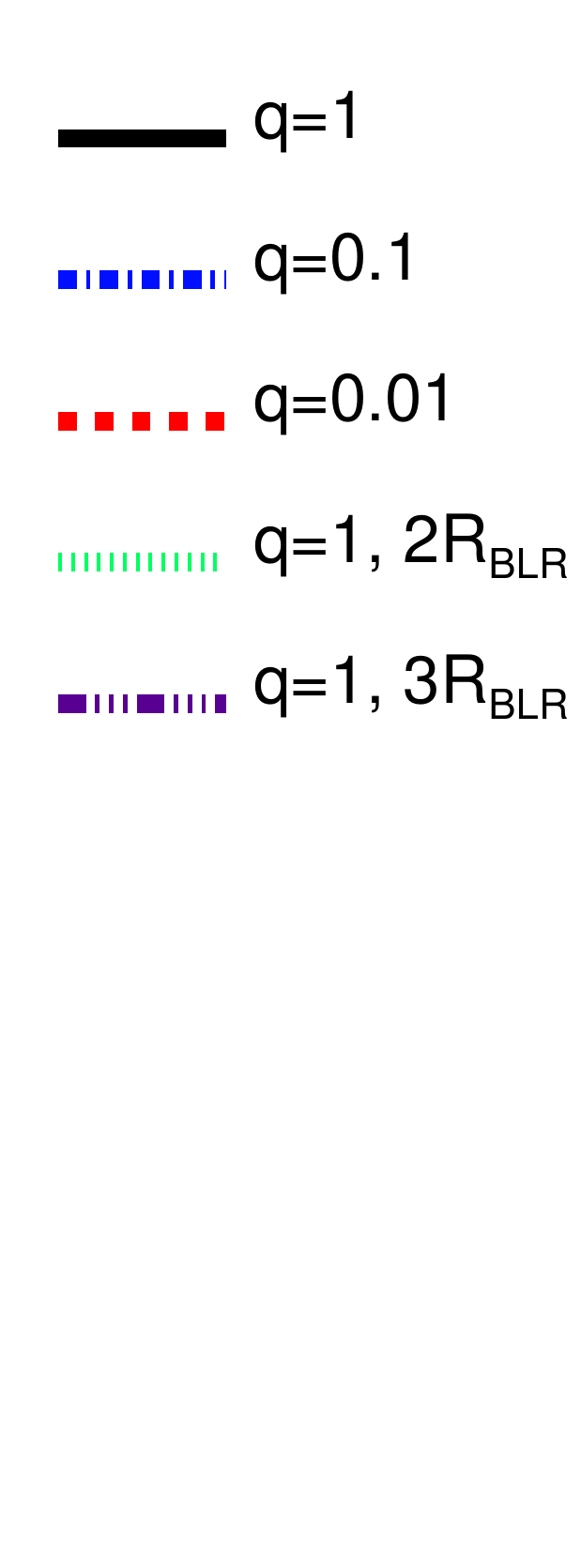}
\caption{
  Observability of SMBH binaries in a next-generation survey assuming
  that at sub-parsec scales, the SMBH binary inspiral is driven by
  interactions with the accretion disks (see \S
  \ref{sec:timeevolution}, Fig. \ref{fig:tres}). Left: $\dot{m}_{\rm
    Edd} = 0.1$. Right: $\dot{m}_{\rm Edd}= 1.0$. The solid (black),
  dash-dotted (blue) and dashed (red) lines are assuming a binary mass
  ratio of 1, 0.1 and 0.01 respectively, while all of them adopt the
  $\rblr$ from \citet{2010shenloeb}. The dotted (green) and
  dash-dotted (purple) lines increase the BLR size by a factor of 2, 3
  respectively, just to show the sensitivity to values of $\rblr$ of
  our results. The big star marks our detection from the SDSS DR7 high
  S/N subsample.  If we reobserve all $10^5$ SDSS quasars with time
  intervals of 10 years, then we would expect $\sim 2 \times 10^4$
  binary detections if all quasars were equal-mass tight binaries, or
  $\sim 4 \times 10^4$ for $q=0.1$ tight binaries.}
\label{fig:fraction_disk}
\end{figure*}

As we discussed in \S \ref{sec:timeevolution}, detectability depends
on the evolutionary scenario of SMBH binaries at sub-parsec scales.
In a scenario where circumbinary disks drive the inspiral of BH
binaries following the $t_{\rm res}$ distribution in Fig. \ref{fig:tres},
the observability as a function of the time interval between
observations of the same quasar is shown in
Fig. \ref{fig:fraction_disk} (left panel for $\dot{m}_{\rm Edd} =
0.1$, and right panel for $\dot{m}_{\rm Edd} = 1.0$) where the star
marks the upper limit on the detectability from our detection in the
SDSS DR7 high S/N subsample.

As before (see \S \ref{sec:results}), the Eddington ratio
$\dot{m}_{\rm Edd}$ makes a big difference because: (1) larger
$\dot{m}_{\rm Edd}$ corresponds to smaller BH masses at a given
luminosity, which leads to smaller velocity drifts at all separations,
and (2) the time $t_{\rm res}$ that BH binaries spend at intermediate
separations is smaller for higher $\dot{m}_{\rm Edd}$. The
distribution of Eddington ratios likely lies between these two limits
for the majority of sources \citep[e.g.][]{2011shen2}. For major
mergers ($q \sim$ 1) with $\dot{m}_{\rm Edd} = 0.1$, we would expect
$\sim$ 20\% of the sources to show orbital motions for a time interval
of $\sim 10$ years, which would correspond to a large number of
targets in next-generation surveys (e.g., 1800 expected candidates
from BOSS QSO sample with multiple observations). Considering our
constraints on the fraction of quasars that are triggered in a tight
binary phase ($<$ 33\%), the expected observability in the scenario
above is $\sim$ 6.7\%. However, this value drops to $\sim$ 1\% if most
sources are closer to $\dot{m}_{\rm Edd} = 1.0$. More extreme mass
ratios would yield much higher observability, given the assumption
that all the observed quasars are the secondary BHs, but to probe such
low mass ratios we likely would need to observe less luminous (lower
mass) QSOs. We should again caution that our predictions are very
sensitive to the assumption of BLR size. As
Fig. \ref{fig:fraction_disk} shows, the observability drops by a
factor of $\sim$ 10 if we increase the BLR size by a factor of 2. If
we increase the BLR size further, e.g. by a factor of 4, we expect
no detections even with a time interval of 100 years (not shown). 
On the contrary, for tidally truncated BLRs, the observability 
would actually increase from our fiducial calculations.

%% f8

\begin{figure}[!t]
\centering
\includegraphics[width=0.5\textwidth]{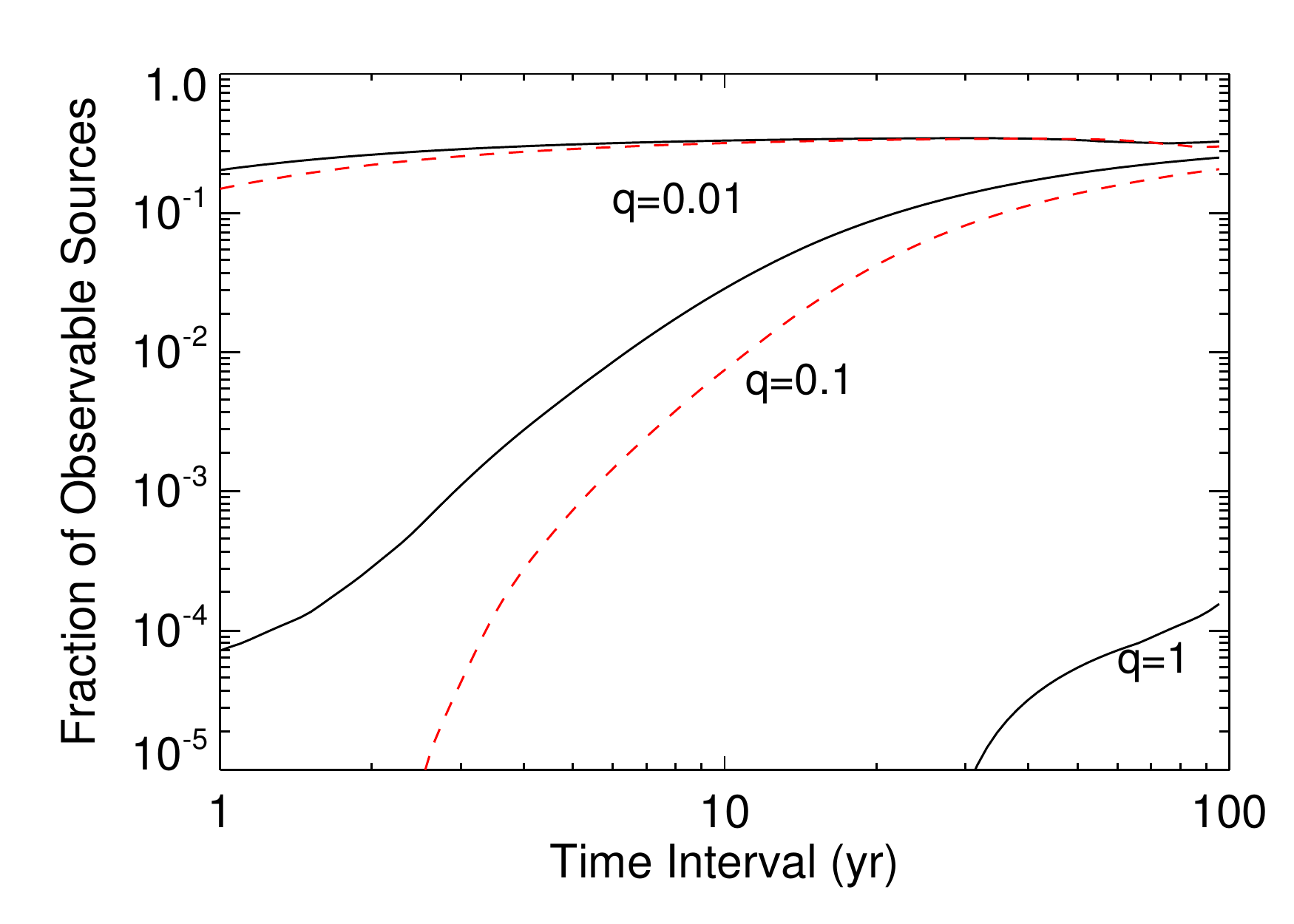}
\caption{ Fraction of observable SMBH binaries in next generation surveys
  assuming that all SMBH binaries stall at sub-parsec scales forever due
  to depletion of stars with no subsequent evolution. The three pairs
  of lines correspond to mass ratio $q = 1$, 0.1, 0.01 from bottom to top
  respectively. The black solid lines utilize black hole masses from
  broad-line width measurements, while the red dashed lines use the
  Eddington-limited BH masses estimated from bolometric luminosity. In the
  $q=1$ case, the stalling radius (Eq. \ref{eq:rstall}) is so large that no BH binaries are expected to be observed within 100 yrs using the Eddington-limited BH mass distribution so there is no red dashed
  line for it.}
\label{fig:fraction_stall}
\end{figure}

In the scenario that the SMBH binaries stall for a Hubble time 
near $\sim 1$ pc, the observability of the binaries
versus time interval between repeated observations is shown in
Fig. \ref{fig:fraction_stall}. Observabilities increase dramatically
as the mass ratio $q$ decreases, because the stalling radius of the
binary due to the depletion of stars is much smaller for smaller
$q$. For comparable-mass binaries, we may only hope to detect a shift in
repeated observations separated by $\sim$ 100
years. Thus, the only way to carve out this parameter space is to push
to lower secondary masses (i.e., lower luminosities than have been
observed by the SDSS).

%% f9

The Subaru Prime Focus Spectrograph (PFS, \citealt{2012ellis}), a new
fiber-fed optical and near-infrared spectrograph that is now being
planned, will be able to study QSOs over a wide redshift range. It
will observe quasars down to $i <$ 24 mag, which is much deeper than
SDSS.  It could also target up to 20 QSOs deg$^{-2}$ from SDSS over
the 1400 deg$^2$ wide survey area. Therefore, nearly all the SDSS QSOs
could be reobserved, providing a very large sample of QSOs with
long-baseline repeat observations.  Considering that the total quasar
sample from DR7 is of order $10^5$ objects, we can think of constraining 
binary fractions as low as $\sim 10^{-4}$, with time baselines greater than a
decade, as potentially observable in the near future.

%%%%%%%%%%%%%%%%%%%%%%%%%%%%%%%%%%%%%%%%%%%%%%%%%%%%%%%%%%%%%%%%%%%%%%%%%%%%%%
%%%%%%%%%%%%%%%%%%%%%%%%%%%%%% Conclusion %%%%%%%%%%%%%%%%%%%%%%%%%%%%%%%%%%%%%%%
%%%%%%%%%%%%%%%%%%%%%%%%%%%%%%%%%%%%%%%%%%%%%%%%%%%%%%%%%%%%%%%%%%%%%%%%%%%%%%
\section{Summary}
\label{sec:summary}

We have employed a new method to search for sub-parsec supermassive
black hole (SMBH) binaries. We make use of repeat observations of
quasars at $0.36 < z < 2$ in the Sloan Digital Sky Survey (SDSS) and
look for temporal changes in the line-of-sight velocity of the
Mg~{\tiny \Rmnum{2}} broad emission line using cross-correlation. 
A radial velocity drift in the broad emission lines can be interpreted as an
orbital acceleration of the accreting black hole and its associated
broad-line region. From the SDSS quasar catalog, we find a sample of
4204 quasars with multiple observations (1791 of them with $\Delta t >
0.5$ year) and select 1523 quasars with S/N per pixel $>$ 10 as a
more reliable subsample. The absolute magnitude in $g$ of the whole
sample ranges from $-18.4$ to $-27.5$ mag with a mean value of $-24.1$
mag. We use two sets of BH masses: Using {\mgii} line widths, the BH
masses range from $10^{7} -10^{10.5}$ $M_\odot$ and
peak at $10^{8.75} M_\odot$; If we simply estimate a lower limit
to the BH mass assuming all sources radiate at 50\% of their Eddington
luminosity we find a mass range of $10^{6.3} - 10^{9.8}$
$M_\odot$ and a peak at $10^{8.3} M_\odot$.

The velocity shifts measured by cross-correlation have a distribution
peaked at 0 km~s$^{-1}$ with $\sigma \sim 100$ km~s$^{-1}$ for the
whole sample and $\sigma \sim 82$ km~s$^{-1}$ for the high S/N
subsample. We set the criteria for our SMBH binary candidates as: the
velocity shift measured with cross-correlation over either the whole
spectral range or only the Mg~{\tiny II} line region should be larger
than $3.4 \sigma$ (280 km~s$^{-1}$), and the difference between the
two measurements should be less than $1~\sigma$ ($\sim$ 82
km~s$^{-1}$). We select seven binary BH candidates out of the high S/N 
subsample, of which three are most reliable. We are now following these up to search for
continued evidence of orbital motion.

Comparing the detectability of SMBH binaries with the number of
candidates ($N \leq 7$), we can rule out that all $10^9$ M$_{\odot}$
BHs exist in $\sim$ 0.03--0.2 pc scale binaries, in a scenario where
binaries stall at a sub-pc scale radius for a Hubble time. In the
scenario of gas-assisted SMBH binary evolution, our analysis reports
that $\lesssim$ 33\% of quasars host SMBH binaries with orbital
separations $<0.1$pc at their centers, subject to outstanding
uncertainties in the BH masses and BLR sizes. 

We should also ask what the cosmological expectations are for such
binaries.  While a full exploration is beyond the scope of this work,
we note from the simulations of \citet{2003volonteri} that massive
host galaxies will have experienced numerous major ($q \gtrsim 0.1$) 
mergers since redshift $z \approx 2$.  Furthermore, according to
simulations, pairing of SMBHs via dynamical friction is most effective
for major mergers (e.g.,
\citealt{2009callegari}).  Thus, our results are already
cosmologically relevant.  In the future, we may hope to constrain the
lifetime of sub-pc binaries with larger samples. 

We also estimate the observability of SMBH binaries with next
generation surveys, such as BOSS or the Prime Focus Spectrograph on
Subaru, in the context of the disk-assisted inspiral scenario.  With
the reobservation of SDSS quasars by these new surveys, we will have a
much larger sample with much longer time baselines to investigate the
temporal change of broad emission lines. These future observations
will help constrain evolutionary scenarios of SMBH binaries.

\acknowledgements We thank an anonymous referee for substantially
improving this work. We thank Michael Strauss for helpful directions
on SDSS data and inspiring discussions. We are also grateful for
helpful comments from Bence Kocsis, Craig Loomis, Frans Pretorius, Alberto Sesana,
Yue Shen, Sean McWilliams, Sarah Burke Spolaor and Justin Ellis.
Wenhua Ju thanks Petchara Pattarakijwanich for useful and patient
discussions on IDL coding.

%%%%%References

%%%%%%%%% Tables %%%%%%%%%%%
\newpage
%% The candidates
\begin{deluxetable}{lllccccccc}
\tablenum{1}
\tablecolumns{10} 
\tabletypesize{\scriptsize}
\tablewidth{0pc}
\tablecaption{Selected Candidates \label{table: selected}}
\tablehead{ 
\colhead{SDSS ID} & \colhead{$z$} & \colhead{M$_{abs}$} & 
\colhead{$\log_{10} M_{\rm BH}$} & \colhead{$\Delta$ t} & \colhead{$\Delta V_1$} & 
\colhead{$\Delta V_2$} & \colhead{Error} & \colhead{$r_{max}$} 
& \colhead{Last Observation}. \\ 
\colhead{} & \colhead{} & \colhead{} & 
\colhead{($M_\odot$)} & \colhead{(yr)} & \colhead{(km~s$^{-1}$)} & 
\colhead{(km~s$^{-1}$)} & \colhead{(km~s$^{-1}$)} & \colhead{(pc)} 
& \colhead{} \\ 
\colhead{(1)} & \colhead{(2)} & \colhead{(3)} & 
\colhead{(4)} & \colhead{(5)} & \colhead{(6)} & 
\colhead{(7)} & \colhead{(8)} & \colhead{(9)} &
\colhead{(10)} 
}
\startdata
%1124  %last MJD 52202
J032223.02+000803.5  &  0.62  & $-$22.8&8.2827	&   0.9123  &   780  &   760  & 0.22 & 0.032 & 10-20-2001\\ 
%0585 %last MJD 52203
%J014136.40+001019.7  &  0.41   &$-$22.2&8.5557	&   1.1370  &  710   &  740   & 0.050 & 10-21-2001\\ 
%0079 %last MJD 51900
J002444.11$-$003221.4*   & 0.40   &	$-$24.3&9.5618	&    0.2301   &  328  &   380 & 2.76 &0.102 & 12-22-2000\\ 
%2211 %last MJD 54530
J095656.42+535023.2$_b$*  &  0.61   &$-$23.1&8.2944	&  6.1589    &   360  &   300  & 1.74 & 0.127 & 03-05-2008\\ 
%3817 %last MJD 52443
J161609.50+434146.8$_m$   & 0.49   &	$-$22.8&8.1696	&  0.2000  &     330   &  250  & 5.10 & 0.021 & 06-18-2002\\ 
%1415 %last MJD 51876
J075700.70+424814.5   & 1.17 &$-$24.7	&9.1311	&   0.0192  &     310    & 290  & 2.12 & 0.020 & 11-28-2000\\ 
%2020 %last MJD 52670
J093502.54+433110.7$_m$    &0.46   &$-$25.6&9.3425	&  0.9452   &     290   &  270   & 2.10 & 0.181 & 01-31-2003\\
%0281 %last MJD 51913
J004918.98+002609.4*    &1.94   &	$-$26.1&9.3148	&  0.2767   &   $-$260   & $-$290 & 5.65 & 0.096 & 01-04-2001\\ 
%0317 
%J005629.32+000109.3   & 1.17    &-24.4&9.0500	& 2.7671e$-$01   &   2.6e+02     &1.8e+02  \\ 
%1519 
%J081403.22+433647.1    &0.44    &-22.1&8.7550	& 6.7945e$-$01  &    -2.6e+02   & -2.1e+02  \\ 
%0148 
%J003714.83+004554.1   & 1.02    &-24.9&8.9260	& 1.2822e+00   &  -2.1e+02   & -2.6e+02  \\ 
\enddata
\tablecomments{
\small
\\
(1) Object name in the SDSS catalog.\\
(2) Redshift of the object from the SDSS catalog \citep{2012bolton}.\\
(3) Absolute $g-$band magnitude.\\
(4) BH mass of the object measured from the Mg~{\tiny \Rmnum{2}} widths \citep{2011shen2}.\\
(5) The time interval between multiple observations.\\
(6) The velocity shift of Mg~{\tiny \Rmnum{2}} measured by cross-correlation over the whole 
wavelength range [1700 \AA, 4000 \AA].\\
(7) The velocity shift of Mg~{\tiny \Rmnum{2}} measured by cross-correlation over the wavelength
range near Mg~{\tiny \Rmnum{2}} [2700 \AA, 2900 \AA]. \\
(8) The uncertainties of velocity shifts reported from peak hunting of cross correlation functions.\\
(9) The upper limit for the separation of the SMBH binary assuming a
mass ratio of $q=1$. The true separations are equal to these values
only when the observer sits at the ideal position for watching
velocity drifts: an edge-on orbit with the
secondary BH just intersecting the line connecting the primary BH
and the observer.  \\
(10) The date when the last SDSS observation of the quasar was taken, in the format of MM-DD-YY.\\
---\\
* The more reliable candidates which passes the test cross-correlation over [1700 \AA, 3800 \AA] (see \S \ref{subsec:uncertainty}).\\
$m$: For these quasars, one of the two epoch spectra has a ``marginal" plate quality flagged in SDSS platelist. These plates get the ``marginal" flag either because the minimum S/N of the all fibers is smaller than 15.0, or the number of exposures is smaller than 3. \\
$b$: One of the two epoch spectra has a ``bad" plate quality flag. This flag reflects the very poor seeing conditions and bright sky when the data were taken, and may lead to large spectrophotometric uncertainty.  However, for this candidate, most of the spurious features are introduced in the red end of the spectrum, far from the Mg~{\tiny II} line.
}
\label{table:candidates}
\end{deluxetable}

\clearpage

\newpage
%% The candidates
\begin{deluxetable}{lllccccccc}
\tablenum{2}
\tablecolumns{9} 
\tabletypesize{\scriptsize}
\tablewidth{0pc}
\tablecaption{Long List of Candidates by {\mgii} Cross-Correlation Alone \label{table: selected2}}
\tablehead{ 
\colhead{SDSS ID} & \colhead{$z$} & \colhead{M$_{abs}$} & 
\colhead{$\log_{10} M_{\rm BH}$} & \colhead{$\Delta$ t} & \colhead{$\Delta V_1$} & 
\colhead{$\Delta V_2$} & \colhead{Error} & \colhead{$r_{max}$} 
& \colhead{Last Observation}. \\ 
\colhead{} & \colhead{} & \colhead{} & 
\colhead{($M_\odot$)} & \colhead{(yr)} & \colhead{(km~s$^{-1}$)} & 
\colhead{(km~s$^{-1}$)} & \colhead{(km~s$^{-1}$)} & \colhead{(pc)} 
& \colhead{} \\ 
\colhead{(1)} & \colhead{(2)} & \colhead{(3)} & 
\colhead{(4)} & \colhead{(5)} & \colhead{(6)} & 
\colhead{(7)} & \colhead{(8)} & \colhead{(9)} &
\colhead{(10)} 
}
\startdata
J104315.10+312313.7 & 1.94 & $-$26.2 & 8.1240 & 0.1288 &          $-$17 &        $-$1863 & 2.71 & 0.006 & 03-01-2005 \\
J004052.14+000057.3 & 0.41 & $-$23.1 & 9.3988 & 1.2822 &         $-$207 &        $-$1501 & 4.89 & 0.097 & 12-18-2001 \\
J013418.19+001536.8 & 0.40 & $-$24.3 & 7.9043 & 1.0493 &          103 &         1050 & 9.98 & 0.019 & 10-21-2001 \\
J212639.96-065555.5 & 1.34 & $-$24.4 & 8.3162 & 0.0603 &            0 &         $-$916 & 1.13 & 0.008 & 10-18-2001 \\
J111045.38+042142.7 & 0.77 & $-$24.2 & 8.1932 & 0.0082 &          $-$17 &         $-$880 & 5.44 & 0.003 & 03-23-2002 \\
J110539.65+041448.3 & 0.44 & $-$23.9 & 8.2472 & 0.0082 &          103 &          862 & 3.59 & 0.003 & 03-23-2002 \\
J232801.46+001705.0$_m$ & 0.41 & $-$22.2 & 8.8313 & 1.0466 &         $-$120 &          845 & 8.03 & 0.061 & 10-18-2001 \\
J120142.24-001639.9 & 1.99 & $-$26.2 & 9.0953 & 0.7315 &          $-$34 &          811 & 4.22 & 0.070 & 01-21-2001 \\
J034512.62+002245.9 & 0.41 & $-$22.2 & 8.8401 & 1.0685 &          103 &          786 & 8.62 & 0.064 & 10-19-2001 \\
J032223.02-000803.5 & 0.62 & $-$22.8 & 8.6694 & 0.9123 &          776 &          759 & 0.22 & 0.050 & 10-20-2001 \\
%J014136.40-001019.7 & 0.41 & $-$22.73 & 8.7042 & 1.1370 &          705 &          742 & 1.14e+03 & 0.058 & 10-21-2001 \\
J002311.06+003517.5 & 0.42 & $-$21.3 & 8.3932 & 0.2301 &           69 &          698 & 7.09 & 0.019 & 12-22-2000 \\
J031830.60+011023.9$_m$ & 0.51 & $-$22.5 & 7.8357 & 1.2740 &           69 &          690 & 1.61 & 0.024 & 01-12-2002 \\
J081931.49+450801.5$_m$ & 1.88 & $-$25.7 & 8.6771 & 0.6794 &          155 &          684 & 7.65 & 0.046 & 10-25-2001 \\
J015454.88+004044.0$_m$ & 1.65 & $-$25.8 & 8.1363 & 0.8986 &          103 &          670 & 2.61 & 0.028 & 10-17-2001 \\
J002444.11+003221.4 & 0.40 & $-$24.3 & 9.3974 & 0.2301 &          604 &          660 & 2.77 & 0.062 & 12-22-2000 \\
J025331.93+001624.8$_m$ & 1.83 & $-$25.3 & 8.7330 & 0.9836 &            0 &          655 & 3.60 & 0.060 & 09-23-2001 \\
J230745.14-004542.7 & 1.84 & $-$25.3 & 8.4820 & 2.9397 &           34 &         $-$638 & 4.91 & 0.078 & 09-02-2003 \\
J025156.31+005706.4$_m$ & 0.47 & $-$22.7 & 8.3595 & 0.8329 &           51 &         $-$621 & 6.99 & 0.037 & 09-23-2001 \\
%J025432.52-004220.1 & 0.43 & $-$23.0 & 8.2728 & 0.9836 &          240 &          621 & 1.57e+03 & 0.036 & 09-23-2001 \\
J135831.39+010339.2 & 0.41 & $-$22.8 & 8.8752 & 0.8247 &           17 &          621 & 2.92 & 0.066 & 02-02-2001 \\
J140140.44+325116.7 & 0.40 & $-$21.6 & 8.4292 & 0.1781 &          $-$34 &          608 & 3.04 & 0.019 & 03-21-2007 \\
J115115.38+003827.0 & 1.88 & $-$26.5 & 8.3378 & 0.7699 &          $-$51 &          604 & 1.78 & 0.035 & 02-03-2001 \\
J105014.02+275230.3 & 1.71 & $-$25.5 & 9.1812 & 0.0712 &         $-$138 &         $-$586 & 5.36 & 0.028 & 04-01-2006 \\
J011310.39-003133.3 & 0.41 & $-$22.4 & 8.3847 & 1.1178 &         $-$172 &          552 & 5.30 & 0.046 & 10-20-2001 \\
J025316.45+010759.8$_m$ & 1.03 & $-$24.3 & 8.7330 & 0.9836 &           17 &         $-$533 & 0.99 & 0.066 & 09-23-2001 \\
J075543.68+421146.0 & 0.42 & $-$22.4 & 8.0433 & 0.0192 &           34 &          519 & 3.19 & 0.004 & 11-28-2000 \\
J105417.14+241326.1 & 1.97 & $-$26.4 & 8.4642 & 0.0493 &           86 &          484 & 7.43 & 0.011 & 12-17-2006 \\
J015638.55+131757.8$_m$ & 1.98 & $-$26.3 & 8.1363 & 0.0630 &          103 &         $-$483 & 9.66 & 0.009 & 12-22-2000 \\
J163905.16+142326.4 & 1.00 & $-$24.0 & 9.3655 & 1.0192 &          276 &          467 & 5.14 & 0.149 & 06-21-2006 \\
J144426.04-010546.0 & 0.56 & $-$22.9 & 9.1813 & 2.0521 &           69 &         $-$465 & 2.17 & 0.171 & 05-17-2002 \\
J082214.83+431702.0$_m$ & 0.97 & $-$23.8 & 8.8360 & 0.6794 &           51 &         $-$455 & 4.81 & 0.067 & 10-25-2001 \\
J145126.16+032643.4 & 0.48 & $-$23.1 & 8.7074 & 0.0438 &          $-$51 &         $-$455 & 3.46 & 0.015 & 05-16-2001 \\
J104913.85+221831.7 & 1.98 & $-$25.6 & 8.6655 & 0.0493 &         $-$224 &         $-$450 & 7.15 & 0.015 & 12-17-2006 \\
J002411.66-004348.1$_m$ & 1.79 & $-$26.2 & 8.7239 & 1.1534 &            0 &          444 & 7.29 & 0.078 & 10-21-2001 \\
%J090823.50+060648.5 & 1.55 & $-$25.3 & 8.8670 & 0.0603 &           53 &          431 & 5.40e+02 & 0.021 & 02-04-2003 \\
J084119.42+171527.5 & 1.75 & $-$25.1 & 8.5174 & 0.0082 &          707 &          428 & 1.95 & 0.005 & 12-07-2005 \\
J151051.34-010906.3 & 1.76 & $-$25.5 & 9.2923 & 2.0247 &           17 &          414 & 2.89 & 0.205 & 05-10-2002 \\
J014209.72-002348.4 & 1.35 & $-$24.4 & 8.7042 & 1.1370 &         $-$759 &         $-$414 & 1.80 & 0.078 & 10-21-2001 \\
J114924.77-010010.3 & 1.13 & $-$24.5 & 8.4314 & 0.7699 &         $-$189 &         $-$396 & 6.27 & 0.048 & 02-03-2001 \\
J033211.85+002430.8 & 0.51 & $-$22.1 & 8.6316 & 0.1890 &           17 &         $-$379 & 4.62 & 0.031 & 12-01-2000 \\
J230654.00-001605.6 & 0.56 & $-$22.9 & 8.8690 & 2.9918 &          $-$69 &          379 & 4.03 & 0.160 & 09-02-2003 \\
J082012.62+431358.4$_m$ & 1.07 & $-$24.8 & 8.5195 & 0.6794 &           86 &          378 & 2.61 & 0.051 & 10-25-2001 \\
J004403.47+151200.0 & 1.78 & $-$25.6 & 9.2250 & 0.1836 &          $-$86 &          374 & 3.91 & 0.060 & 12-01-2000 \\
J003451.86-011125.7 & 1.84 & $-$25.6 & 8.8775 & 1.2822 &          $-$34 &         $-$362 & 4.28 & 0.108 & 12-18-2001 \\
J020646.97+001800.6 & 1.68 & $-$25.5 & 9.0745 & 0.1781 &          $-$51 &          362 & 2.78 & 0.051 & 11-29-2000 \\
J091621.10+012020.8$_m$ & 2.00 & $-$26.3 & 9.2819 & 0.0712 &           69 &         $-$362 & 2.99 & 0.041 & 02-15-2001 \\
J163359.76-004512.6 & 1.24 & $-$24.2 & 8.8806 & 0.0685 &           69 &         $-$349 & 3.73 & 0.026 & 06-01-2000 \\
J014415.12-002349.8 & 1.71 & $-$25.4 & 9.5618 & 1.1370 &          $-$51 &          345 & 0.91 & 0.230 & 10-21-2001 \\
J124502.76+380514.2 & 1.05 & $-$24.9 & 8.8668 & 0.7589 &           86 &         $-$345 & 5.57 & 0.084 & 02-06-2006 \\
J150638.54+573459.1 & 0.56 & $-$23.8 & 8.6990 & 0.0630 &            0 &         $-$343 & 1.44 & 0.020 & 06-19-2001 \\
%J111059.90+053947.3 & 0.98 & $-$23.9 & 8.1932 & 0.0082 &         -226 &         -343 & 5.40e+02 & 0.004 & 03-23-2002 \\
J164535.42+362712.3 & 1.81 & $-$26.0 & 8.0411 & 1.0219 &           34 &         $-$340 & 8.91 & 0.038 & 05-23-2003 \\
J080915.74+273830.3$_m$ & 1.80 & $-$25.5 & 8.4768 & 0.1425 &          $-$86 &         -337 & 7.88 & 0.024 & 01-31-2003 \\
J161052.15+442421.3 & 1.23 & $-$24.4 & 8.1199 & 0.2000 &           51 &         $-$334 & 1.21 & 0.019 & 06-18-2002 \\
%J131401.18-031342.9 & 1.30 & $-$24.6 & 7.9140 & 0.8192 &          -23 &          330 & 7.09e+02 & 0.030 & 03-22-2001 \\
J153111.83+442354.6 & 1.14 & $-$24.7 & 8.3325 & 0.0055 &           69 &         $-$329 & 8.81 & 0.004 & 07-13-2002 \\
J025257.18-010220.9 & 1.25 & $-$24.2 & 8.7452 & 0.1671 &          $-$17 &          318 & 2.47 & 0.036 & 11-29-2000 \\
J002028.35-002915.0 & 1.93 & $-$25.3 & 8.3291 & 0.2301 &         $-$224 &          310 & 3.10 & 0.026 & 12-22-2000 \\
%J095145.24+525305.8 & 1.79 & $-$25.7 & 8.4035 & 6.1589 &          125 &         -310 & 4.45e+02 & 0.149 & 03-05-2008 \\
J163523.71+134036.6 & 1.81 & $-$25.3 & 8.7050 & 0.0822 &           17 &          310 & 2.06 & 0.024 & 06-21-2006 \\
J215213.51-074224.9$_m$ & 0.95 & $-$24.3 & 8.3290 & 0.0658 &            0 &          310 & 6.56 & 0.014 & 09-21-2001 \\
J102536.43+645242.9 & 1.37 & $-$24.9 & 8.3168 & 0.0438 &          224 &          307 & 3.21 & 0.011 & 01-21-2001 \\
J095656.42+535023.2$_m$ & 0.61 & $-$23.1 & 7.9160 & 6.1589 &          362 &          303 & 1.74 & 0.086 & 03-05-2008 \\
J014905.16+005925.3 & 1.00 & $-$24.2 & 8.5702 & 1.1205 &          $-$69 &         $-$299 & 5.93 & 0.078 & 10-20-2001 \\
J155631.51+080051.8$_m$ & 1.92 & $-$25.8 & 8.6527 & 1.7288 &          241 &         -294 & 4.00 & 0.108 & 05-04-2006 \\
J020527.77-005747.6 & 1.24 & $-$24.7 & 9.0239 & 0.1781 &          $-$86 &          293 & 5.66 & 0.053 & 11-29-2000 \\
J075700.70+424814.5 & 1.17 & $-$24.7 & 8.9236 & 0.0192 &          310 &          293 & 2.12 & 0.016 & 11-28-2000 \\
J085626.71+570444.7 & 1.52 & $-$24.6 & 8.9563 & 0.1151 &            0 &         $-$293 & 7.97 & 0.039 & 02-02-2001 \\
J004918.98-002609.4 & 1.95 & $-$26.1 & 7.9173 & 0.2767 &         $-$258 &         $-$289 & 5.65 & 0.019 & 01-04-2001 \\
%J093503.77+003739.1 & 1.09 & $-$24.6 & 8.9033 & 0.7863 &           34 &          282 & 2.24e+03 & 0.099 & 02-09-2002 \\
\enddata
\tablecomments{
\tiny
(1) Object name in the SDSS catalog.
(2) Redshift of the object from the SDSS catalog \citep{2012bolton}.
(3) Absolute $g-$band magnitude.
(4) BH mass of the object measured from the Mg~{\tiny \Rmnum{2}} widths \citep{2011shen2}.
(5) The time interval between multiple observations.
(6) The velocity shift of Mg~{\tiny \Rmnum{2}} measured by cross-correlation over the whole 
wavelength range [1700 \AA, 4000 \AA].
(7) The velocity shift of Mg~{\tiny \Rmnum{2}} measured by cross-correlation over the wavelength
range near Mg~{\tiny \Rmnum{2}} [2700 \AA, 2900 \AA]. 
(8) The uncertainties of velocity shifts reported from peak hunting of cross correlation functions.
(9) The upper limit for the separation of the SMBH binary assuming a
mass ratio of $q=1$. The true separations are equal to these values
only when the observer sits at the ideal position for watching
velocity drifts: an edge-on orbit with the
secondary BH just intersecting the line connecting the primary BH
and the observer. 
(10) The date when the last SDSS observation of the quasar was taken, in the format of MM-DD-YY. \\
---\\
$m$: For these quasars, one of the two epoch spectra has a ``marginal" plate quality flagged in SDSS platelist. These plates get the ``marginal" flag either because the minimum S/N of the all fibers is smaller than 15.0, or the number of exposures is smaller than 3.
}
\label{table:candidates-large}
\end{deluxetable}

\clearpage

%% sensitivity of results to parameters
\begin{deluxetable}{lcc}
\tablenum{2}
\tablecolumns{3} 
\tabletypesize{\scriptsize}
\tablewidth{0pc}
\tablecaption{Expected number of observable sources in S/N/pixel $>10$ subsample.}
\tablehead{ 
\colhead{} & \colhead{$\dot{m}_{\rm Edd}=0.1$} & \colhead{$\dot{m}_{\rm Edd}=1.0$}
}
\startdata
q=1.0 & 24  & 0.08\\
q=0.1 & 209 & 44\\
\enddata
\tablecomments{
  The four sets of numbers show the expected number of observable
  sources in our S/N/pixel $>10$ subsample, adopting four sets of ($q,
  \dot{m}_{\rm Edd}$) in the scenario of gas-assisted sub-pc evolution
  of the SMBH binaries. We take the virial BH masses for the $\dot{m}_{\rm Edd}=0.1$ cases 
and Eddington-limited BH masses for the $\dot{m}_{\rm Edd}=1$ cases. For comparison, 
in the no-evolution case with $r=0.1$ pc, we expect 16-48 sources for $q=1$ and 
336-479 sources for the unrealistic $q=0.1$ case.}
\label{table:results}
\end{deluxetable}

\end{document}